\documentstyle[lprocl,psfig,11pt]{article}

\def\ar#1#2#3{Ann.\ Rev.\ Nucl.\ Part.\ Sci.\ #1 (19#3) #2}

\def\np#1#2#3{Nucl.\ Phys.\ B#1 (19#3) #2}
\def\pl#1#2#3{Phys.\ Lett.\ #1B (19#3) #2}
\def\pr#1#2#3{Phys.\ Rev.\ D #1 (19#3) #2}
\def\prep#1#2#3{Phys.\ Rep.\ #1 (19#3) #2}
\def\prl#1#2#3{Phys.\ Rev.\ Lett.\ #1 (19#3) #2}

\def\sj#1#2#3{Sov.\ J.\ Nucl.\ Phys.\ #1 (19#3) #2}
\def\zp#1#2#3{Z.\ Phys.\ C#1 (19#3) #2}
\def\acta#1#2#3{Acta\ Phys.\ Polon.\ B#1 (19#3) #2}
\def\ar#1#2#3{Annu.\ Rev.\ Nucl.\ Part.\ Sci.\ #1 (19#3) #2}

\def\beq{\begin{equation}}
\def\eeq{\end{equation}}
\def\beeq{\begin{eqnarray}}
\def\eeeq{\end{eqnarray}}
\def\ee{e^+e^-}
\def\as{\alpha_S}
\def\to{\rightarrow}
\def\ltap{\raisebox{-.4ex}{\rlap{$\,\sim\,$}} \raisebox{.4ex}{$\,<\,$}}
\def\gtap{\raisebox{-.4ex}{\rlap{$\,\sim\,$}} \raisebox{.4ex}{$\,>\,$}}
\def\ycut{y_{\rm cut}}
\def\st{\scriptstyle}
\def\ho{{\rm \st higher \; orders}}
\def\higher{\raisebox{-.7ex}{\rlap{$\ho$}}
\raisebox{.4ex}{$\;\;\; \;\;\longrightarrow$}}
\def\a0{\alpha_{0}}
\def\ab0{{\overline {\a0}}}
\def\amz{\alpha_s(M_{\rm Z^0})}
\def\epem{e^+e^-}
\def\z0{\rm Z^0}

\begin{document}

\title{QCD AT HIGH ENERGIES}

\author{STEFANO CATANI}

\address{INFN, Sezione di Firenze and 
         Dipartimento di Fisica,
         Universit\'a di Firenze, \\
         Largo E. Fermi 2, I-50125  Florence, Italy \\
         and \\
         TH Division, CERN, 
         CH-1211 Geneva 23, Switzerland \\
         E-mail: catani@vxcern.cern.ch}
        
%


\begin{titlepage}
\renewcommand{\thefootnote}{\fnsymbol{footnote}}
\begin{flushright}
     CERN-TH/97-371 \\ hep-ph/9712442
     \end{flushright}
\par \vspace{10mm}
\begin{center}
{\Large \bf
QCD AT HIGH ENERGIES\footnote{Invited talk given at the XVIII International 
Symposium on Lepton Photon Interactions, LP97, Hamburg, Germany, 
July 28th--August 1st, 1997. To be published in the Proceedings.}}  
\end{center}
\par \vspace{2mm}
\begin{center}
{\bf Stefano Catani}\\

\vspace{5mm}

{I.N.F.N., Sezione di Firenze and Dipartimento
di Fisica, Universit\`a di Firenze}\\
{Largo E. Fermi 2, I-50125 Florence, Italy}\\
{and}\\
{Theory Division, CERN}\\
{CH-1211 Geneva 23, Switzerland}
\end{center}

\par \vspace{2cm}
\begin{center} {\large \bf Abstract} \end{center}
\begin{quote}
The following topics in perturbative QCD are reviewed: recent theoretical
progress in higher-order calculations; soft-gluon resummation for 
hard-scattering processes at large $E_T$ and high $x$; low-$x$ behaviour of
structure functions and recent theoretical results on BFKL dynamics;
infrared renormalons and power corrections to
perturbative predictions; updated summary of
$\as$ measurements.
\end{quote}
\vspace*{\fill}
\begin{flushleft}
     CERN-TH/97-371 \\ December 1997
\end{flushleft}
\end{titlepage}

\maketitle\abstracts{
The following topics in perturbative QCD are reviewed: recent theoretical
progress in higher-order calculations; soft-gluon resummation for 
hard-scattering processes at large $E_T$ and high $x$; low-$x$ behaviour of
structure functions and recent theoretical results on BFKL dynamics;
infrared renormalons and power corrections to
perturbative predictions; updated summary of
$\as$ measurements.}

\section{Introduction}
\label{intro}

Quantum Chromodynamics (QCD) is nowadays 
settled as the theory of strong 
interactions within the Standard Model (SM) of the elementary particles.
To a large extent, 
this achievement is the result of the theoretical and experimental
progress in hadronic physics at high energies. In this kinematic
regime, QCD is synonymous of `perturbative QCD'. 
This talk will cover
a selection of topics, rather than results, in perturbative QCD. 
More comprehensive reviews of recent results on jet physics~\cite{Schel},
unpolarized~\cite{Schek} and polarized~\cite{brull} nucleon structure functions,
photon structure~\cite{SSR}, diffractive processes~\cite{gallo} and 
heavy-quark~\cite{HQtalk} production and decay
are presented in other contributions to this Symposium.

Calculations at the lowest order (LO) in QCD perturbation theory
give only the order of magnitude of hard-scattering cross-sections. The
theoretical accuracy of perturbative-QCD predictions is instead controlled by 
the size of the next-to-leading order (NLO) and, in general, higher-order 
contributions. Section~\ref{HOcal} is devoted to a review of the
theoretical progress in perturbative calculations and also includes the results
of recent measurements of $\as$. 

Soft-gluon radiation is a source of large higher-order corrections for   
hard-scattering processes near the exclusive boundary of the phase space. 
In these cases, summation of the corrections to all orders can be important
to improve the accuracy of the perturbative approach. 
In Sect.~\ref{sgresumm}, some predictions based 
on soft-gluon resummation are briefly reviewed 
and their impact on the analysis of experimental data from high-energy colliders
is discussed.

Low-$x$ physics is a topic at the border of hard and soft hadronic interactions.
Section~\ref{lxp} summarizes present analyses of the low-$x$ behaviour of the 
nucleon structure function and outlines recent theoretical developments 
in the BFKL formulation of small-$x$ dynamics.

At high energies, non-perturbative phenomena affect perturbative predictions by
contributions that are suppressed by inverse powers of the energy.
Some recent theoretical ideas and phenomenological studies to quantify
power corrections are discussed in Sect.~\ref{irpc}.

A world summary of $\as$ determinations is presented in Sect.~\ref{sumas}
and some concluding remarks are left to Sect.~\ref{conc}.

\section{Higher-order calculations in perturbation theory}
\label{HOcal}

The evaluation of perturbative corrections to hard-scattering cross-sections
requires the computation of higher-order Feynman diagrams that involve real and 
virtual partons. In this computation one 
has to deal with different kinds of singularities. The customary ultraviolet 
singularities, present in the virtual contributions, are removed by 
renormalization. By adding real and virtual terms, the infrared divergences
cancel in inclusive cross sections, while the left-over collinear 
singularities are factorized in the process-independent 
parton distributions and fragmentation functions.

Because of this complicated pattern of singularities, it is natural to 
divide QCD observables into two different classes, according to their degree of 
inclusiveness.

\subsection{Completely inclusive quantities: NNLO predictions and recent
$\as$ determinations}
\label{nnloas}

Fully inclusive quantities are infrared- and collinear-safe observables
that depend on a single momentum scale. They can be 
expressed as a simple power series expansion in $\as$.

The best known observables of this
type are the total hadronic cross-section in $\ee$ annihilation and the
hadronic branching ratio of the $Z^0$. In Ref.~\cite{Ce},
these observables were computed
up to next-to-next-to-leading order (NNLO) in perturbation theory,
i.e. to relative accuracy ${\cal O}(\as^3)$ with respect to the lowest-order
approximation. These calculations have recently been confirmed by an 
independent re-evaluation~\cite{Chetyrkin}. Using these results, one can  
parametrize~\cite{Passarino} the hadronic branching ratio of the $Z^0$
as follows
\beq
\label{rz}
R_Z=\frac{\Gamma_{\rm had}(M_Z)}{\Gamma_{\rm lep}(M_Z)} \simeq R_0 
\left[1+1.06 \,\frac{\as}{\pi} 
+ 0.9 
\left(\frac{\as}{\pi}\right)^2-15 \left( \frac{\as}{\pi}\right)^3 \right] 
+ {\cal O}\left(\left(\frac{1}{M_Z}\right)^p \,\right) \;, 
\eeq
where the factor $R_0$ includes the electroweak radiative corrections
and, in particular, depends on the values of the masses of the top quark and
Higgs boson. The term ${\cal O}((1/M_Z)^p)$ stands for 
non-perturbative power corrections (cf. Sect.~\ref{irpc}).

Other quantities, which have been computed at NNLO,
are the hadronic width of the $\tau$ lepton~\cite{pich,neubert} and
the deep inelastic scattering (DIS) sum rules~\cite{sr}, 
namely the Gross-Llewellyn Smith sum rule~\cite{GLSsr} and the 
polarized~\cite{Bjsr} and unpolarized Bjorken sum rules. 

Fully inclusive observables are the simplest quantities that can accurately be
evaluated in QCD perturbation theory. Their (relative) simplicity
from the computational viewpoint follows from kinematics. Since these 
observables are completely inclusive, no phase-space restriction has to be 
applied. Real and virtual contributions can be combined at the 
integrand level and this produces the cancellation of infrared and collinear
singularities before performing the relevant phase-space integrations. Owing to
these features, general techniques have been
available for some time~\cite{russi} 
to carry out NNLO calculations in analytic form. These techniques are suitable 
for automatic implementations in computer codes~\cite{NNLOaut}.

The high accuracy of the theoretical predictions for these observables
derives from the availability of NNLO calculations and from the fact that
non-perturbative power corrections can be controlled by operator product
expansions~\cite{ope} (OPEs) and are relatively small (in general,
$p \geq 2$ and, in particular, $p=4$ in Eq.~(\ref{rz})). Because of these 
reasons, fully inclusive observables are particularly suitable for 
$\as$ determination.

The cleanest and, in principle, most accurate determination of $\as$ is 
that performed at the $Z^0$ peak. Here, effects of power corrections are
strongly suppressed because of the large value of the $Z$ mass, and the
theoretical uncertainty due to perturbative-QCD contributions beyond
NNLO is estimated to be $(\Delta \as)_{\small QCD} = \pm 0.002$. In practice,
however, the value of $\as$ obtained in this way is quite sensitive
to the assumption that electroweak interactions are accurately described
by the SM. For instance, extracting $\as(M_Z)$ from  
Eq.~(\ref{rz}), one has~\cite{haidt}
\beq
\label{dasew}
\as = (\as)_{\small S.M.} -
3.2 \; 
\frac{\delta\Gamma_{\rm had}}{(\Gamma_{\rm had})_{\small S.M.}} \;\;.
\eeq
One can see that a variation of the hadronic width by one per mille with respect
to the value expected from the SM produces an effect on $\as$ that is larger
than the QCD uncertainty.

Using a fixed 
top mass, $m_t=175.6\pm 5.5 {\rm GeV}$,
and the experimental value of $R_Z$ reported at the 1997 summer 
conferences~\cite{ewsc}, one obtains from Eq.~(\ref{rz})
\beq
\label{asrz}
\as(M_Z)=0.124 \pm 0.004 \,({\rm exp.}) \pm 0.002 \,(m_H) \;\;,
\eeq
where the central value corresponds to the Higgs mass $m_H=300 \, {\rm GeV}$
and the second error is due to the variation of $m_H$ in the range
$60\,{\rm GeV} < m_H < 1\,{\rm TeV}$. The sensitivity of $\as$ to the SM 
assumption can be reduced by considering a global fit to electroweak data.
In this case $\as$ mainly depends on $R_Z$, the total width $\Gamma_Z$ and
the peak value of the hadronic cross section $\sigma_h^0$.
The simultaneous fit of $m_t, \as, m_H$ gives~\cite{ewsc}
$m_t=173.1\pm 5.4 \,{\rm GeV}$ and
\beq
\label{asew}
\as(M_Z)=0.120 \pm 0.003 \,({\rm exp.}) \;, 
\;\;\;m_H = 115^{+ 116}_{- \; \; 66} \,{\rm GeV} \;.
\eeq
The difference between the values (\ref{asrz}) and (\ref{asew}) for $\as$ comes
from a shift ($\sim - 0.002$) produced by the new entry $\Gamma_Z$ and
a further shift ($\sim - 0.002$) due to the different central value of $m_H$.
This shows that, within a global SM fit, $\as$ is still quite
sensitive to $m_H$. This dependence can be parametrized as follows~\cite{haidt}
\beq
\label{asvmh}
(\Delta \as)_{m_H} = 0.0023 \;x_H (1 + 0.2 \;x_H) \;, 
\;\;\;x_H \equiv \ln (m_H/100 {\rm GeV}) \;.
\eeq

Much theoretical work~\cite{masscor} has recently been devoted to the 
calculation of quark-mass corrections to current--current correlators. These
corrections are important for measurements of $\as$ 
from low-energy $\ee$ data.

These results and estimates of non-perturbative 
contributions (along the lines of similar analyses for the hadronic decay of 
the $\tau$ lepton~\cite{pich}) have been used~\cite{jamin} to determine
$\as$ from the $\ee$ hadronic cross-section at the $\Upsilon$ resonance. This
NLO determination gives
\beq
\label{asups}
\as(4.1{\rm GeV})= 0.228^{+ 0.045}_{-0.030} \;, 
\eeq
where the error is dominated by the theoretical uncertainty.
The corresponding value evolved to the $Z^0$ mass
is $\as(M_Z)=0.119^{+0.010}_{-0.008}$ and has to be
compared with a previous determination~\cite{Vol}, $\as(M_Z)=0.109 \pm 0.001$,
based on LO predictions.

Measurements of $\as$ from the continuum in $\ee$ annihilation at low energies
have typically large experimental errors due to poor statistics.
The new measurement submitted by the CLEO Collaboration to this Symposium
is~\cite{rcleo}
\beq
\label{ascleo}
\as(10.52 \,{\rm GeV})= 0.20 \pm 0.01 ({\rm stat.}) \pm 0.06 ({\rm syst.}) \;. 
\eeq
Note that, because of the large amount of data collected 
at CESR just below the $\Upsilon$ resonance, the statistical error is no longer 
dominant. The systematic error is theoretical (uncertainties in QED 
radiative corrections) and experimental (estimates of backgrounds and detector
efficiencies).

\subsection{Inclusive quantities: NLO calculations and general algorithms}

QCD calculations beyond LO for inclusive quantities are much more 
involved. Owing to the complicated phase space for multiparton configurations,
analytic calculations are in practice impossible for most of the distributions.
Moreover, infrared and collinear singularities, separately present in the real
and virtual contributions at the intermediate 
steps, have to be first regularized by analytic continuation in a number of 
space-time dimensions $d=4-2\epsilon$ different from four. This analytic 
continuation prevents a straightforward implementation of numerical 
integration techniques. 

Methods to overcome these problems are known.
They consist in using hybrid analytical/numerical procedures:
one must somehow extract and simplify the singular parts and treat them
analytically; the remainder is treated numerically, independently of the full 
complications of the actual calculation. These methods were first~\cite{ERT} 
used to evaluate 3-jet cross sections in $\ee$ annihilation and  
were then applied to other cross sections~\cite{book}, 
adapting the method each time to the particular process. This is very 
time-consuming and has required lot of effort to produce new NLO calculations.

Only recently has it become clear that these methods are generalizable in a 
process-independent manner. 
The key observation is that the singular parts of the QCD matrix
elements can be singled out in a general way by using
the factorization properties of soft and collinear radiation.
Owing to this universality, the methods have led to general 
algorithms~\cite{GGjet,FKSjet,CSdipole} for NLO QCD calculations.  

The various algorithms use different methods and techniques (phase-space
slicing~\cite{GGjet}, subtraction method~\cite{FKSjet}, 
dipole formalism~\cite{CSdipole}) to achieve a common goal. All the analytical 
work that is necessary to evaluate and cancel the infrared singularities is 
carried out once and for all.
The final output of the algorithms is given in terms
of effective matrix elements that can be automatically constructed starting
from the original (process-dependent) matrix elements and universal 
(process-independent) factors. The effective matrix elements
can be integrated numerically or analytically (whenever possible)
over the available phase space in four dimensions to compute the actual value 
of the NLO cross section. If the numerical approach is chosen, Monte Carlo 
integration techniques can be easily implemented to provide general-purpose 
Monte Carlo programs for carrying out NLO 
calculations in any given process.
 
Using these algorithms, the computation of inclusive quantities at NLO 
essentially amounts to the evaluation of the original matrix elements.
Since efficient techniques~\cite{mangano} (based on 
helicity formalism and colour-subamplitude decompositions) are available
for calculating real matrix elements, the
computation of the matrix elements for the virtual contribution 
remains the only real obstacle to perform new NLO calculations.

In recent years, this obstacle was greatly reduced by the introduction
of new tools~\cite{BDKrev}, inspired by string-theory methods, for the
evaluation of one-loop amplitudes. One-loop matrix elements involving
up to five massless partons are known~\cite{5part} and the computation
of those with four
massless partons and a vector boson has been completed recently~\cite{V4part}.

The relevance of the theoretical progress that I have briefly outlined
is witnessed by the accelerated production rate of new NLO calculations.
These include, for instance, 4-jet cross sections~\cite{4ee} and mass
quark corrections~\cite{3massive} to 3-jet observables in $\ee$ 
annihilation. The complete calculation for 3-jet cross sections in
hadron collisions~\cite{3ppbar} is expected to appear soon.
 
The theoretical accuracy of NLO predictions for inclusive observables
can reach the level of 10--20$\%$, although in some notable cases 
(e.g. production cross sections of charm and bottom quarks~\cite{HQtalk}
and direct photons~\cite{Schel}) 
theoretical uncertainties are much larger and
the agreement with data is poor.
This accuracy is very important
for physics studies within and beyond the SM. The comparison between
NLO predictions and high-energy collider data is discussed in 
Ref.~\cite{Schel} and in the following sections. The new calculations
mentioned above are extremely valuable for further studies such as,
for instance, to obtain stronger constraints on the elusive light 
gluino~\cite{4ee,gluino} and to measure the running of the $b$-quark 
mass~\cite{3massive,delphimb}.

The extension to higher orders of the NLO techniques described in this section
is still a challenge for theorists. The high experimental accuracy of
LEP and SLC data demands improved perturbative-QCD
predictions and efforts in this direction.

\vspace*{-1mm}
\section{All-order resummation, high $E_T$, high $x$}
\label{sgresumm}

Higher-order perturbative contributions are certainly important for studies
of QCD observables close to the exclusive boundary of the phase space.
Although infrared and collinear singularities cancel in inclusive cross sections
upon adding real and virtual contributions, in this kinematic regime
real emission is strongly inhibited.
The ensuing mismatch of real and virtual corrections generates 
double-logarithmic terms of the type $(\as L^2)^n$, where $L = \ln 1/y$, and
$y$ generically denotes the distance from the exclusive boundary. For instance,
$y=\ycut$ can be the resolution parameter for the transverse size of jets in
$\ee$ annihilation,
or $y=1-2E_T/{\sqrt S}$, where $E_T$ is the transverse energy of jets produced
in hadron collisions at the centre-of-mass energy $\sqrt S$, or $y=1-x$, where
$x$ is the Bjorken variable in DIS processes.

In all these processes, when $y \ll 1$ the logarithmically-enhanced terms of 
infrared origin spoil the convergence of the fixed-order 
expansion in $\as$. Accurate predictions require the evaluation and
(whenever feasible) the resummation of this class of contributions to
all orders in perturbation theory~\cite{softrev}. In the following I briefly
discuss some examples of soft-gluon resummation in lepton and hadron 
collisions. Transverse-momentum distributions of vector bosons and 
di-photon systems are reviewed elsewhere~\cite{Schel} in these proceedings.

\subsection{Jet rates and event shapes in $\ee$ annihilation}
\label{jetratessub}

A detailed understanding of logarithmically-enhanced terms 
exists for jet rates~\cite{CDOTW} and for many shape 
variables~\cite{thrust,hjm} in $\ee$ annihilation in the two-jet limit. In
these cases all-order resummation takes an exponentiated form. For instance,
using the $k_{\perp}$-algorithm~\cite{CDOTW,jetalg} to define jets, the 2-jet 
rate $R_2$ is given by~\cite{CDOTW,diss} 
\beq
\label{r2res}
R_2(\as(Q),\ycut) \equiv \frac{\sigma_{2jet}}{\sigma_{tot}} =
\exp \left\{ - \frac{2C_F}{\pi} \int_{Q{\sqrt {\ycut}}}^Q \frac{dk}{k} \;\as(k)
\ln \frac{Q^2}{k^2} + \dots
\right\} \;\;,
\eeq
where $\ycut$ is the jet resolution parameter, $Q$ is the $\ee$
centre-of-mass energy and the dots denote subleading contributions. Note
that the exponent involves a double-logarithmically-weighted integral of the
QCD coupling. The effective scale $k$ at which $\as$ is evaluated corresponds
to the transverse momentum exchanged at the elementary QCD vertices rather than
to the total momentum $Q$. Using the QCD running, 
$\as(k)=\as(Q)/(1+2 \beta_0 \as(Q) \ln k/Q)$, one can nonetheless perform the
integral in Eq.~(\ref{r2res}) and explicitly obtain $R_2$ as a function of
$\as(Q)$ and $L=\ln 1/\ycut$.   

In general, shape variable and jet rates can written as follows~\cite{CTTW}
\beeq
\label{RCSig}
R(\as,y) = C(\as)\Sigma(\as,L) &+& D(\as,y)\; , \\
\label{lnr}
\Sigma (\as,y) = 
\exp \left\{ L g_1(\as L) + g_2(\as L) + \cdots \right\} &\;,& \;\;\; 
C(\as) = 1 + \sum_{n=1}^\infty C_n \as^n \;,
\eeeq
where $D(\as,y)$ vanishes as $y\to 0$ order-by-order in perturbation theory. 
In the expression
(\ref{RCSig}) the singular $\ln y$-dependence is entirely included in the
effective form factor $\Sigma$, given in Eq.~(\ref{lnr}).
The function $L \,g_1$
resums all the {\em leading} contributions $\as^n L^{n+1}$, while $g_2$
contains the {\em next-to-leading} logarithmic terms $\as^n L^n$ and so forth.
Equation~(\ref{lnr}) represents an improved 
perturbative expansion in the two-jet region $y \ll 1$. 
Once the functions $g_i$ have 
been computed, one has a systematic perturbative
treatment of the shape distribution throughout the region of $y$ in which 
$\as L\ltap 1$, which is much larger than the domain $\as L^2 \ll 1$ in which
the fixed-order expansion in $\as$ is applicable.
Furthermore, the resummed expressions (\ref{RCSig}) and (\ref{lnr}) can  
consistently be matched with fixed-order calculations. 
In particular, one can consider the 
next-to-leading logarithmic approximation (NLLA) as given by the 
functions $g_1$ and $g_2$ and 
combine them with the complete NLO results~\cite{ERT,event} 
(after subtracting the resummed logarithmic terms in order to avoid double 
counting), to obtain a prediction
(${\cal O}(\as^2)$+NLLA) which is everywhere at least as good as the
fixed-order result, and much better as $y$ becomes small.

Detailed experimental studies~\cite{LEPresum} performed at LEP 1 and the SLC 
proved that resummed predictions substantially improve the QCD description
of the data. These analyses led to determinations of $\as$ from hadronic final
states with reduced theoretical uncertainty with respect to 
pure NLO calculations. The combined value from SLD and the four LEP experiments 
is $\as(M_Z)= 0.122 \pm 0.001 ({\rm exp.}) \pm  0.006 ({\rm th.})$.

\begin{figure}
  \centerline{
    \setlength{\unitlength}{1cm}
    \begin{picture}(0,5)
       \put(0,0){\includegraphics{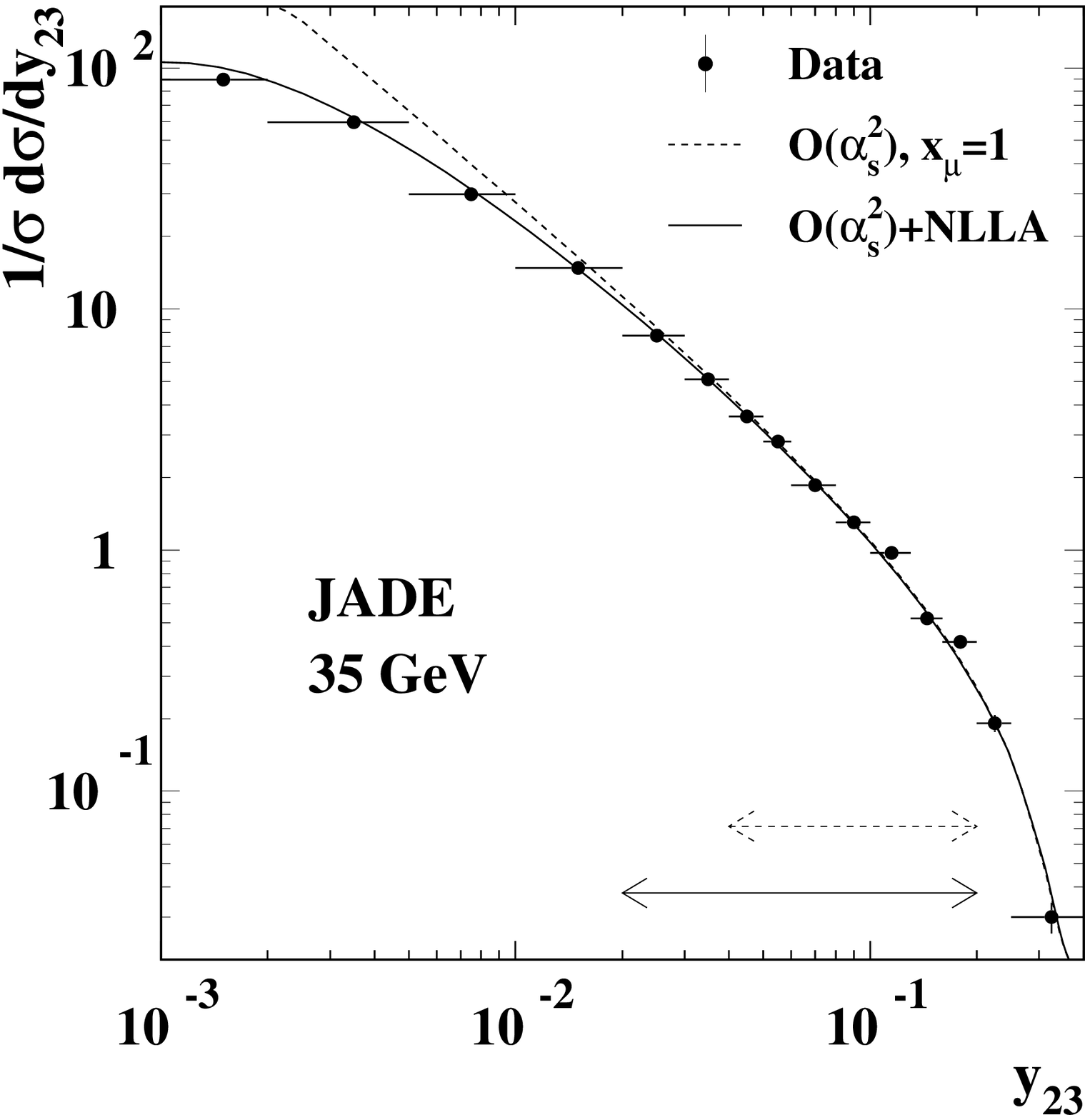}}
       \put(5,0){\includegraphics{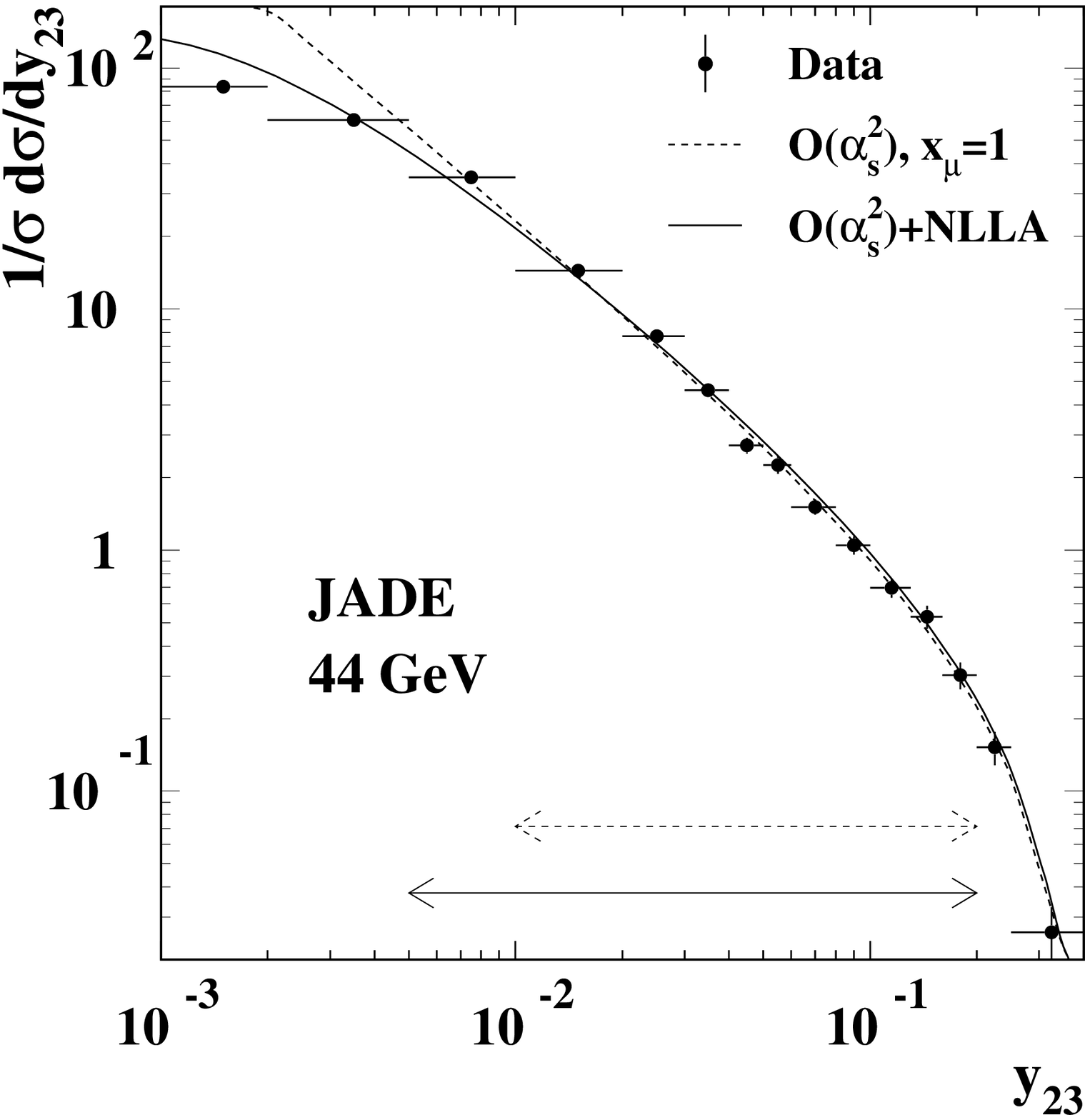}}
    \end{picture}}
\caption{The differential 2-jet rates, $dR_2/dy_{23} 
\;(y_{23}\equiv \ycut)$, measured at $Q=44,35 \;{\rm GeV}$, are shown after
correction to the parton level. The solid and dashed lines correspond 
to the results of fits with resummed and fixed-order calculations, respectively.
Note the extension of the fit range (denoted by the arrow) towards the 
small-$y_{23}$ region in the case of resummed predictions.
\label{figjade}}
\end{figure}

Resummed calculations were also compared with $\ee$ data at lower energies,
namely at PEP~\cite{pep} ($Q=29\, {\rm GeV}$) and TRISTAN~\cite{tristan}
($Q=58\, {\rm GeV}$). A re-analysis of the data recorded in 1981--1986 by the
JADE detector at PETRA has been presented recently~\cite{jaderev}. This 
analysis provides measurements of $\as$ at three different centre-of-mass
energies, $\as(22\,{\rm GeV})=0.161, \;\as(35\,{\rm GeV})=0.143, 
\as(44\,{\rm GeV})=0.137$. The energy dependence of these values is in 
agreement with the QCD expectation and corresponds to
\beq
\label{asjade}
\as(M_Z) = 0.122^{+ 0.008}_{-0.006} \;,
\eeq
where the error is dominated by the theoretical uncertainty. The result 
(\ref{asjade}) improves a previous determination, $\as(M_Z)=0.119 \pm 0.014$,
based on NLO predictions. The improvement is due to the use of more observables
(see, e.g., Fig.~\ref{figjade}) and to the inclusion of more detailed 
systematic studies.

\begin{figure}
  \centerline{
    \setlength{\unitlength}{1cm}
    \begin{picture}(0,5.2)
       \put(0,0){\includegraphics{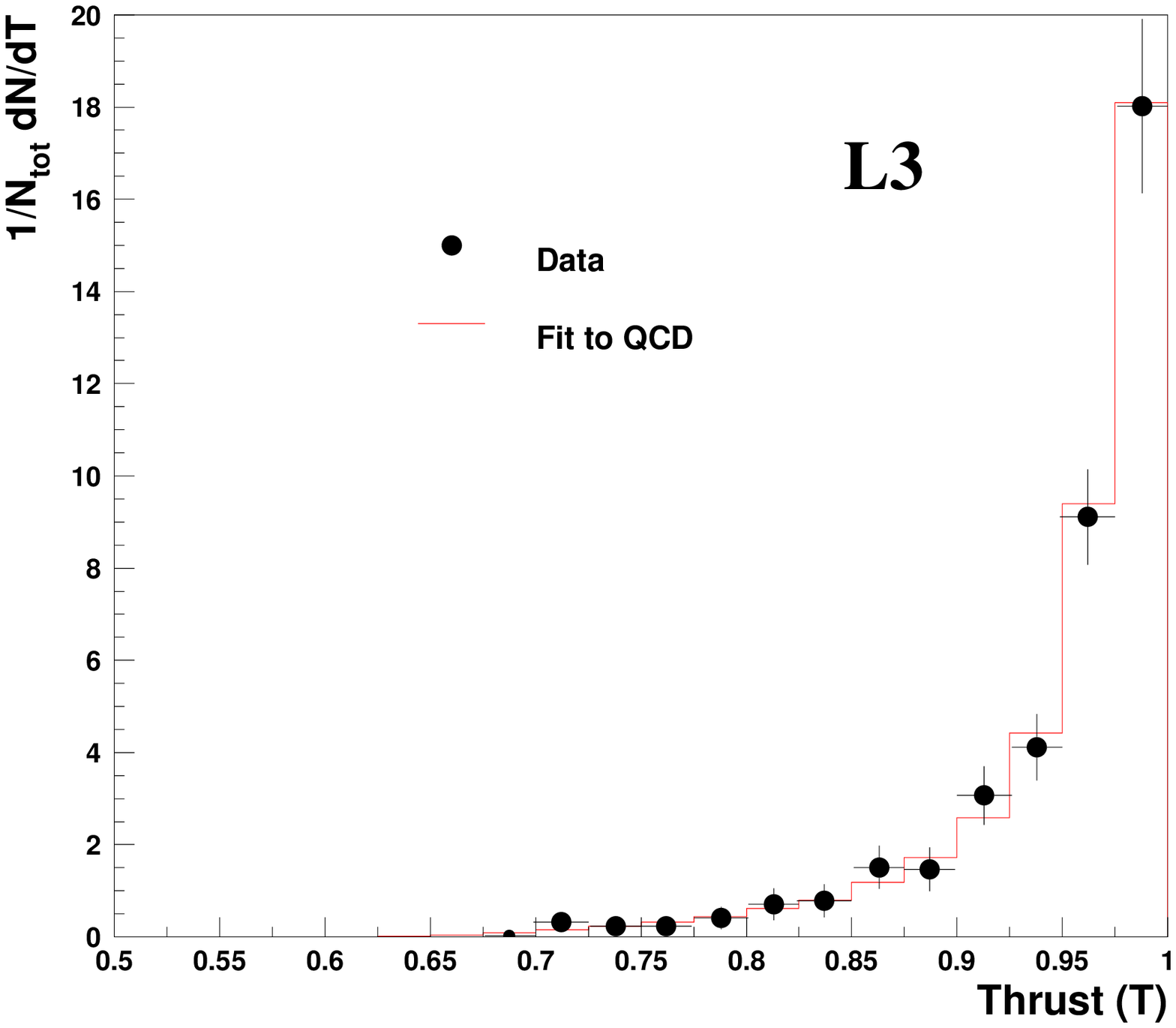}}
       \put(3,0){\includegraphics{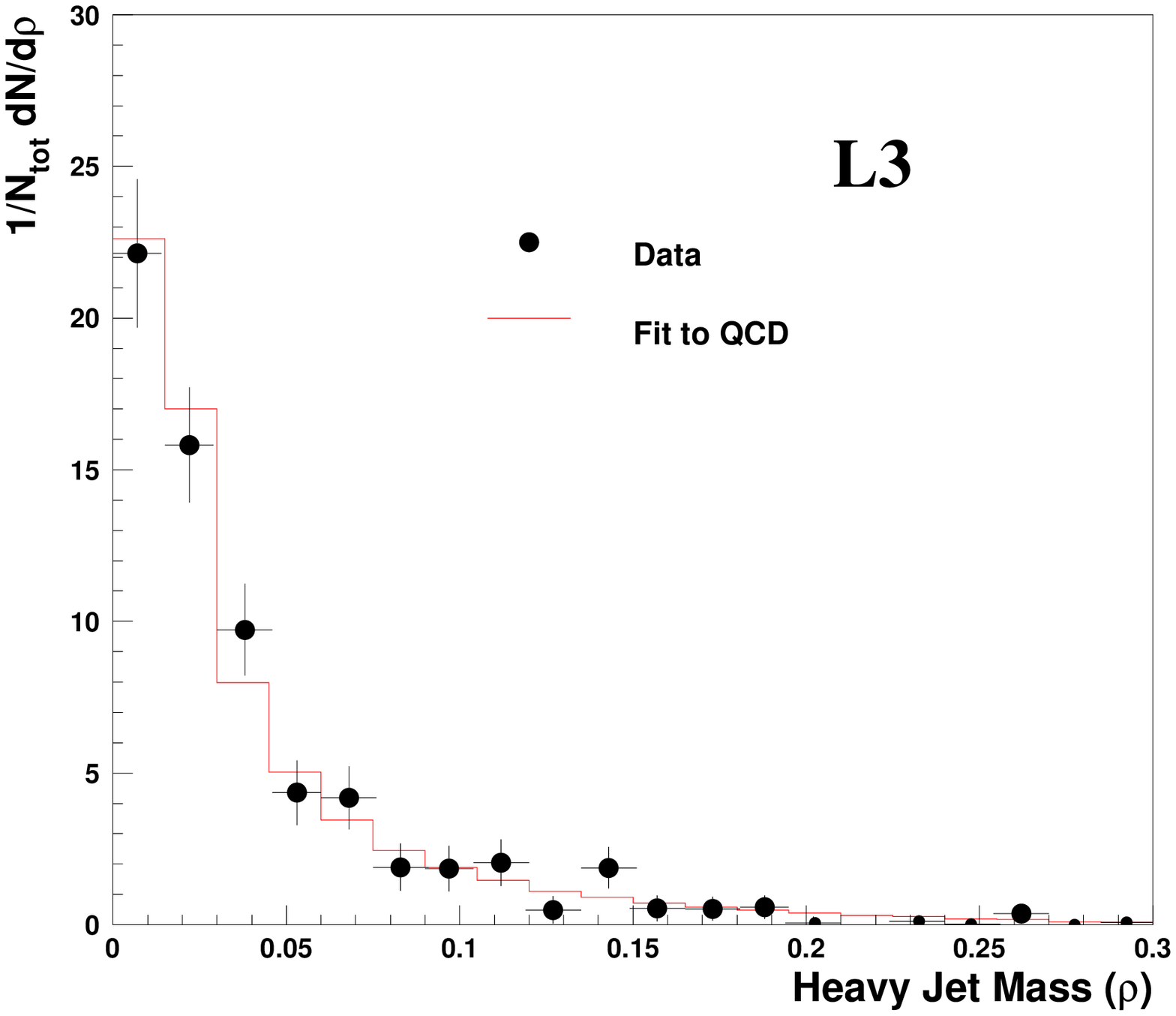}}
    \end{picture}}
\vspace*{-0.6cm}
\caption{Measured distributions of thrust $T$ (left) and scaled heavy jet mass
$\rho$ (right) in comparison with ${\cal O}(\as^2)$+NLLA QCD predictions
at $Q=172 \;{\rm GeV}$.
\label{figl3}}
\end{figure}

The most recent data~\cite{lep2} on hadronic events from LEP 2 around $Q=133, 
161$ and 172~GeV are well described by QCD predictions. Resummed calculations
are quite valuable because they can be used 
for studies of hadronic observables close to the 2-jet region 
(Fig.~\ref{figl3}),
thus increasing the (otherwise very limited) statistics. 
The ensuing first measurements of $\as$ at these new energies are summarized
in Sect.~\ref{sumas}. Assuming the QCD running, the combined LEP 2 average gives
$\as(M_Z)=0.115 \pm 0.008$.

\subsection{Hadron collisions at high $E_T, Q^2$ and $x$}

At high-energy hadron colliders, the region close to the kinematic boundary
of the phase space is the most sensitive to possible 
signals~\cite{cdfjet,heraq2} of new physics. Reliable QCD predictions are thus
important not only to test the SM but also to correctly estimate SM backgrounds.
Note that these predictions depend on an extra non-perturbative input 
with respect to $\ee$ collisions: the parton densities of the incoming
hadrons to be convoluted with hard-scattering cross-sections. The present
knowledge of the parton densities is reviewed elsewhere~\cite{Schek} in these
proceedings.

The top quark is produced relatively close to threshold at the
Tevatron~\cite{top}. 
Calculations of its production cross-section, based on NLO QCD~\cite{HQnlo}
and including soft-gluon resummation, were performed in 
Refs.~\cite{sigres}$^-$\cite{CMNT}. The results disagree on
soft-gluon effects. The disagreement does not regard the general 
framework~\cite{Sterman,kidon} to carry out the resummation, but  
the proper way to implement resummed formulae in actual 
computations of hadronic cross-sections. 

The differences between 
Refs.~\cite{sigres,Berger,CMNTtop} do not have a large impact on
top-quark phenomenology: the various numerical results are consistent 
within uncertainty estimates and no sizeable reduction of the
present experimental error is expected. However, the use of
different approaches~\cite{sigres,Berger,CMNT} to soft-gluon resummation in 
hadron collisions can be more relevant in other processes
as, for instance, in the case of jet production at large 
transverse energy $E_T$.

\begin{figure}
  \centerline{
    \setlength{\unitlength}{1cm}
    \begin{picture}(0,6)
       \put(0,0){\includegraphics{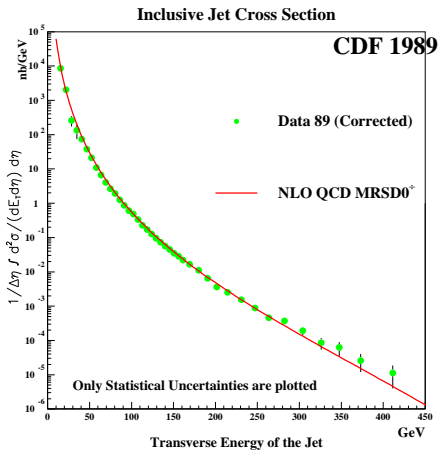}}
    \end{picture}}
\vspace*{-0.1mm}    
\caption{One-jet inclusive cross-section in
$p$$\bar p$ collisions at 
1.8~TeV.
\label{figjetxs}}
\end{figure}

\begin{figure}
  \centerline{
    \setlength{\unitlength}{1cm}
    \begin{picture}(0,7)
       \put(0,0){\includegraphics{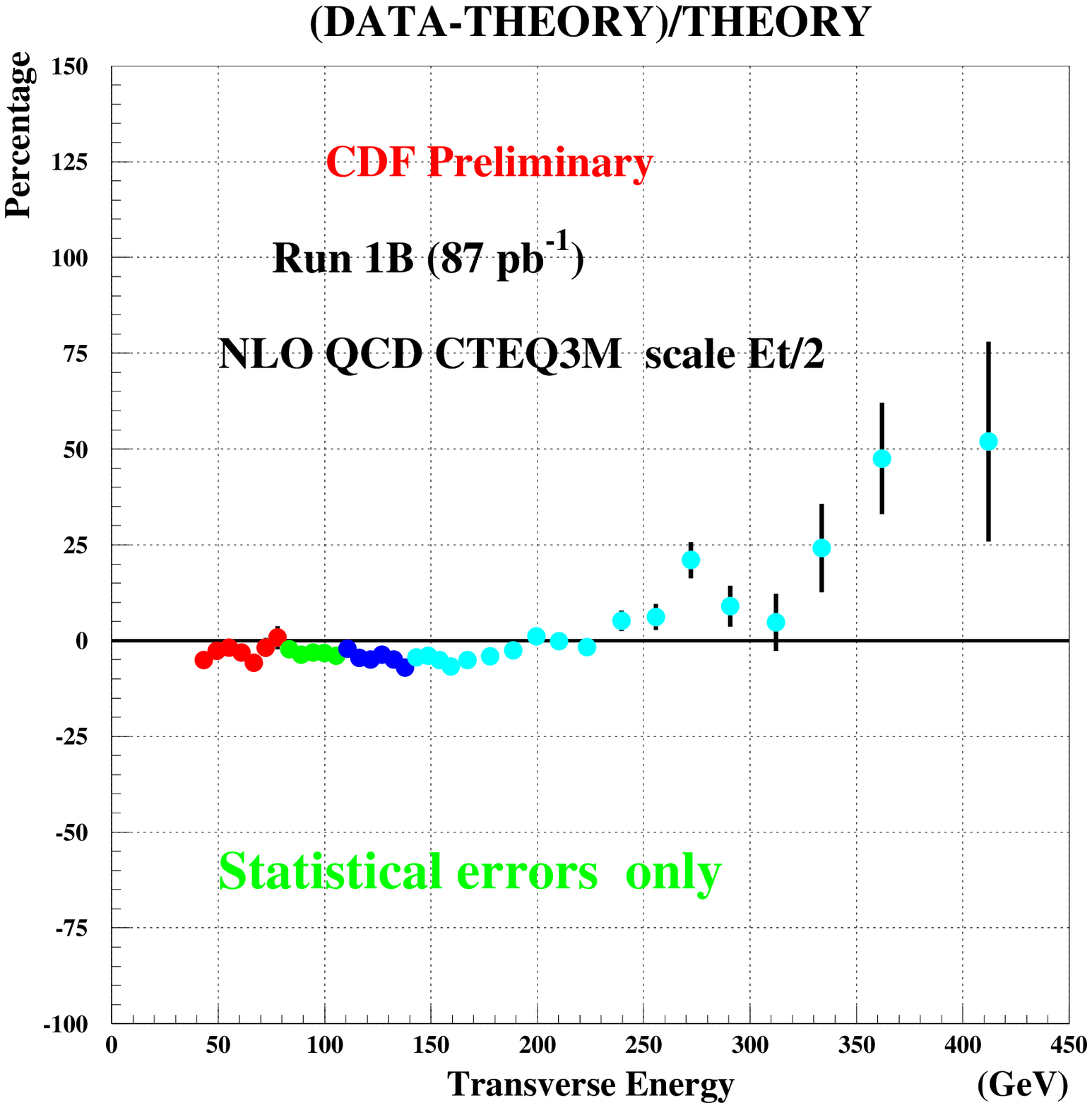}}
       \put(7.5,0){\includegraphics{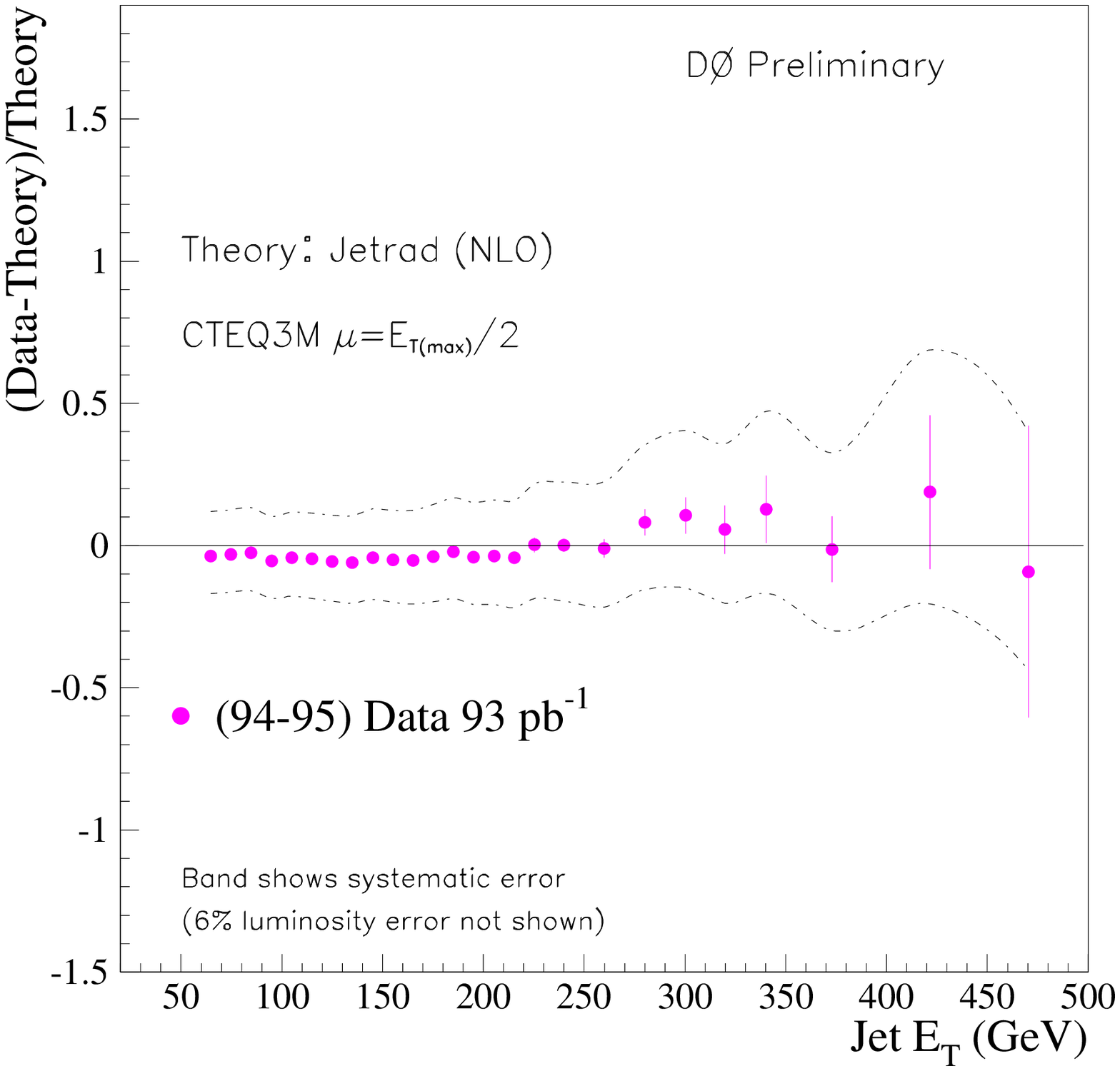}}
    \end{picture}}
\vspace*{-3mm}
\caption{Relative difference between the inclusive jet 
cross-section and NLO predictions. CDF (left) and D0 (right) data refer to 
the pseudorapidity ranges $0.1 < |\eta| < 0.7$ and $|\eta| < 0.5$, respectively.
CDF systematic errors are not shown.
\label{figcdfd0}}
\end{figure}

\begin{figure}
  \centerline{
    \setlength{\unitlength}{1cm}
    \begin{picture}(0,6.5)
       \put(0,0){\includegraphics{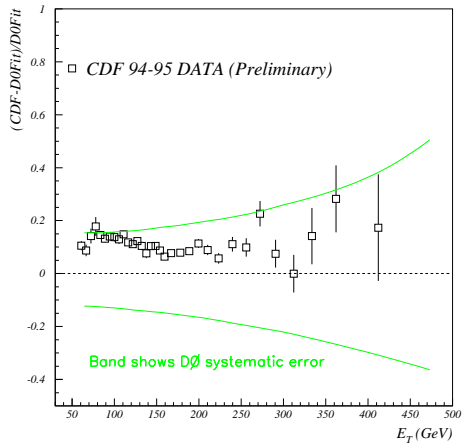}}
    \end{picture}}
\vspace*{-4mm}
\caption{Relative difference between CDF data and
a fit of the D0 data in the range $0.1 < |\eta| < 0.7$. The systematic band
is mainly due to the D0 jet energy scale uncertainty.
\label{figcomp}}
\end{figure}

The one-jet inclusive cross-section measured at the Tevatron~\cite{Schel}
impressively agrees
with NLO QCD predictions~\cite{NLOjet} over almost 9 orders of magnitude
(Fig.~\ref{figjetxs}). However, the CDF Collaboration reported~\cite{cdfjet} an
excess of events at high $E_T \;(E_T \gtap 250 \,{\rm GeV})$. The present 
situation~\cite{Schel,tevjet} can be summarized as follows 
(Fig.~\ref{figcdfd0}):
CDF still sees an excess while D0 does not 
 and, nonetheless, CDF and D0 data are consistent
within experimental errors (Fig.~\ref{figcomp}). Previous data from 
the two experiments could not be compared directly as they probe different
regions of jet pseudorapidity $\eta$. A new D0 analysis in the same
pseudorapidity range as that of CDF shows (Fig.~\ref{figcomp}) that the CDF 
data lie above the D0 results, but within the D0 uncertainty band.

The origin of the differences between these experimental results has to be
further clarified. The angular distribution for two-jet events~\cite{2jets}
is in good agreement with NLO QCD calculations, suggesting that the possible 
high-$E_T$ excess may have an explanation within the framework of QCD rather
than originating from new physics. Independently, the accuracy of the 
QCD predictions has to be carefully estimated.

Using different conventional sets~\cite{grv,mrs,cteq} of parton distributions
and different values of factorization/renormalization scales, NLO
predictions for the one-jet cross-section in the range $50 \ltap E_T \ltap 400$ 
vary by approximately $15\%$. As for soft-gluon effects, one
can at present estimate an increase of the cross section by 
$\sim 20\%$~\cite{Bergernew} or by less than $10\%$~\cite{CMNT}, depending
on the resummation approch used. Further investigations are warranted. 
The fact that the estimate in Ref.~\cite{CMNT} is within the 
variation of the NLO predictions may suggest
that higher-order QCD corrections are under control.

Note, however, that the true uncertainty due to parton densities can be 
larger. As shown by the CTEQ Collaboration~\cite{cteq4hj}, including the CDF
data in global fits to parton distributions, there is enough flexibility 
to increase the gluon density at large $x$ and enhance by 25--30$\%$
the NLO predictions for the single-jet distribution at high $E_T$.

The HERA experiments have reported~\cite{heraq2} an excess of DIS events at
large values of $Q^2\; (15\,000 \ltap Q^2 \ltap 50\,000 \,{\rm GeV}^2)$ 
and $x\; (0.5 \ltap x \ltap 0.7)$. This has raised considerable interest in the 
high-energy physics community~\cite{gual}.
The measured excess has decreased as the integrated
luminosity increased~\cite{straub}, but it still deserves 
attention~\cite{gual}.  

The observed excess is with respect to the SM expectation based on QCD 
predictions. Their accuracy relies on estimates of beyond-NLO perturbative 
corrections and on accurate determinations of non-perturbative
parton densities. 

\begin{figure}
  \centerline{
    \setlength{\unitlength}{1cm}
    \begin{picture}(0,5)
       \put(0,0){\includegraphics{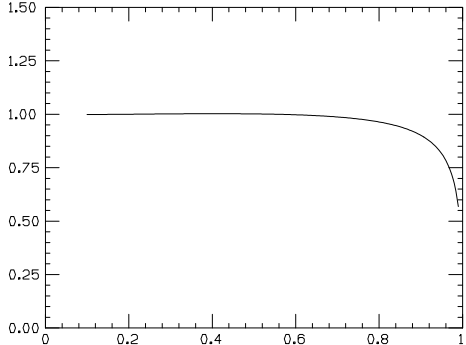}}
    \end{picture}}
\caption{The ratio 
$F^{{\rm res}}(x,Q^2)$$/F^{{\rm NLO}}(x,Q^2)$ 
at $Q^2=30\,000$~GeV$^2$. The 
DIS structure functions $F^{{\rm NLO}}$ and 
$F^{{\rm res}}$ are respectively
obtained by $Q^2$-evolution in NLO and including soft-gluon resummation.
The $Q^2$-evolution is performed by starting from the scale
$Q_0^2=10$~GeV$^2$ with an input structure function
$F(x,Q_0^2) \sim (1-x)^3$.
\label{fighx}}
\end{figure}

In DIS processes at large $x$, logarithmically-enhanced contributions due to
soft-gluon emission can be sizeable. We have implemented~\cite{lpexer} the
corresponding resummation formulae~\cite{CMW} to estimate the order of 
magnitude of the effect in the relevant HERA region. The results of this 
excercise are illustrated in Fig.~\ref{fighx}. The overall sign of the 
resummation
effect on the measured structure functions is negative because multigluon
radiation increases the scaling violation at large $x$. The decrease in the
predictions is nonetheless extremely small for $x \ltap 0.7$, so that
the NLO evolution is safe in this kinematics region.

The main source of uncertainty in the SM predictions~\cite{heraq2} 
thus comes from the extraction of the parton (quark) densities from 
low-$Q^2$ data. From a dedicated fit~\cite{highxunc} to SLAC, BCDMS, NMC 
data at large $x$, the uncertainty is estimated to be $\sim 10\%$. 
This value is larger than the differences ($\sim 6\%$) obtained
by calculations that use different sets~\cite{grv,mrs,cteq} 
of conventional parton densities
(cf. Fig.~3 in Ref.~\cite{highxunc}). Sizeable additional effects due, 
for instance, to a new component~\cite{tung} in the quark densities at 
extremely large values of $x$, can be excluded~\cite{gual} by combining HERA
data in the neutral- and charged-current channels. 
    
\vspace*{-1mm}
\section{Low-$x$ physics and structure functions}
\label{lxp}

Strong-interaction dynamics in the small-$x$ 
regime~\cite{sxrev} is a main challenge to QCD.
Much progress in the field has been prompted by measurements of 
stucture functions~\cite{Schek} and diffractive interactions~\cite{gallo}
at HERA in a kinematic range that extends down to $x \sim 10^{-4}$.
The present status of small-$x$ physics for hard-scattering processes
involving two transverse-momentum scales (e.g. forward jet production in DIS,
two-jet inclusive mesurements at large rapidity gaps) 
is reviewed in Ref.~\cite{Schel}. 

I shall consider single-scale processes
and, in particular, DIS structure functions at low $x$. In this case, one can
study~\cite{Schek} the transition between the perturbative and non-perturbative
regimes, and this may eventually lead to a QCD understanding of   
soft hadronic physics at high energy.

\vspace*{-1mm}
\subsection{Confronting DGLAP with BFKL}
\vspace*{-1mm}
The QCD analysis of low-$x$ structure functions has to deal with a
non-trivial interplay between
perturbative evolution towards the hard scale $Q^2$ and non-perturbative 
behaviour of the parton densities. This complicates 
the answer to a basic question
that regards the region of applicability of the QCD parton model. How far 
in $x$ can we safely use the DGLAP evolution equations~\cite{AP}
before they have to be supplemented with BFKL-type~\cite{BFKL} effects?

To discuss this point let me recall the main steps in the QCD analysis of the 
structure functions. The measurement of the proton structure function
$F_2(x,Q^2) \sim f_S(x,Q^2)$ directly determines the sea-quark density
$f_S= x (q + {\bar q})$. Then one can use the DGLAP evolution equations
(the symbol $\otimes$ denotes the convolution integral with respect to $x$), 
\beeq
\label{df2}
dF_2(x,Q^2) / d\ln Q^2 
&\sim& P_{qq} \otimes f_S +  P_{qg} \otimes f_g \;\;, \\
\label{dfg}
df_g(x,Q^2)/ d\ln Q^2 
&\sim& P_{gq} \otimes f_S +  P_{gg} \otimes f_g \;\;, 
\eeeq
to extract a gluon density $f_g(x,Q^2)= x g(x,Q^2)$ that agrees with 
the measured scaling violation in $dF_2(x,Q^2) / d\ln Q^2$ 
(according to Eq.~(\ref{df2}))
and fulfils the self-consistency equation (\ref{dfg}). 

The perturbative-QCD 
ingredients in this analysis are the Altarelli--Parisi splitting functions
$P_{ab}(\as(Q),x)$. They are computable as power-series expansions in $\as$ and 
are known up to NLO accuracy. The truncation of the splitting functions
at a fixed perturbative order is equivalent to assuming that the dominant
dynamical mechanism leading to scaling violations is the evolution of
parton cascades with strongly-ordered transverse momenta. However, at high 
energy this evolution takes place over large rapidity intervals $(\Delta  y \sim
\ln1/x)$ and diffusion in transverse momentum becomes relevant. Formally, this
implies that higher-order corrections to $P_{ab}(\as,x)$ are logarithmically
enhanced:
\beq
\label{plog}
P_{ab}(\as,x) \sim \frac{\as}{x} + \frac{\as}{x} \;( \as \ln x ) + \dots
+ \frac{\as}{x} \;( \as \ln x )^n + \dots \;\;.
\eeq
At asymptotically small values of $x$, resummation of these corrections is 
mandatory to obtain reliable predictions.

Small-$x$ resummation is, in general, accomplished by the BFKL 
equation~\cite{BFKL}, whose structure is completely known only to 
leading logarithmic (LL) accuracy. In the context of structure-function
calculations, the BFKL equation provides us with resummed expressions for
the splitting functions. 

In the small-$x$ region the gluon channel dominates. Considering the fixed-order
expansion of the splitting function, one-gluon exchange gives 
$P_{gg}(\as,x) \sim \as/x$.
Then, assuming a flat $x$-behaviour of $f_g$ at 
low momentum scales, the evolution equation (\ref{dfg}) produces a gluon
density $f_g \sim \exp ({\sqrt {\ln 1/x}})$ that steeply increases in $1/x$
at higher momentum scales~\cite{first}. The steep behaviour drives strong
scaling violations in $F_2$ and leads to structure functions that strongly
rise as $x$ decreases. This prediction is usually referred to as DGLAP 
prediction.

The increase of the gluon density is 
steeper after BFKL resummation. Summing the LL terms in $P_{gg}$, one 
eventually gets the power-like behaviour $f_g \sim x^{-\lambda_g}$, 
where $\lambda_g \simeq 2.65 \,\as$. 

\begin{figure}
  \centerline{
    \setlength{\unitlength}{1cm}
    \begin{picture}(0,14)
       \put(0,0){\includegraphics{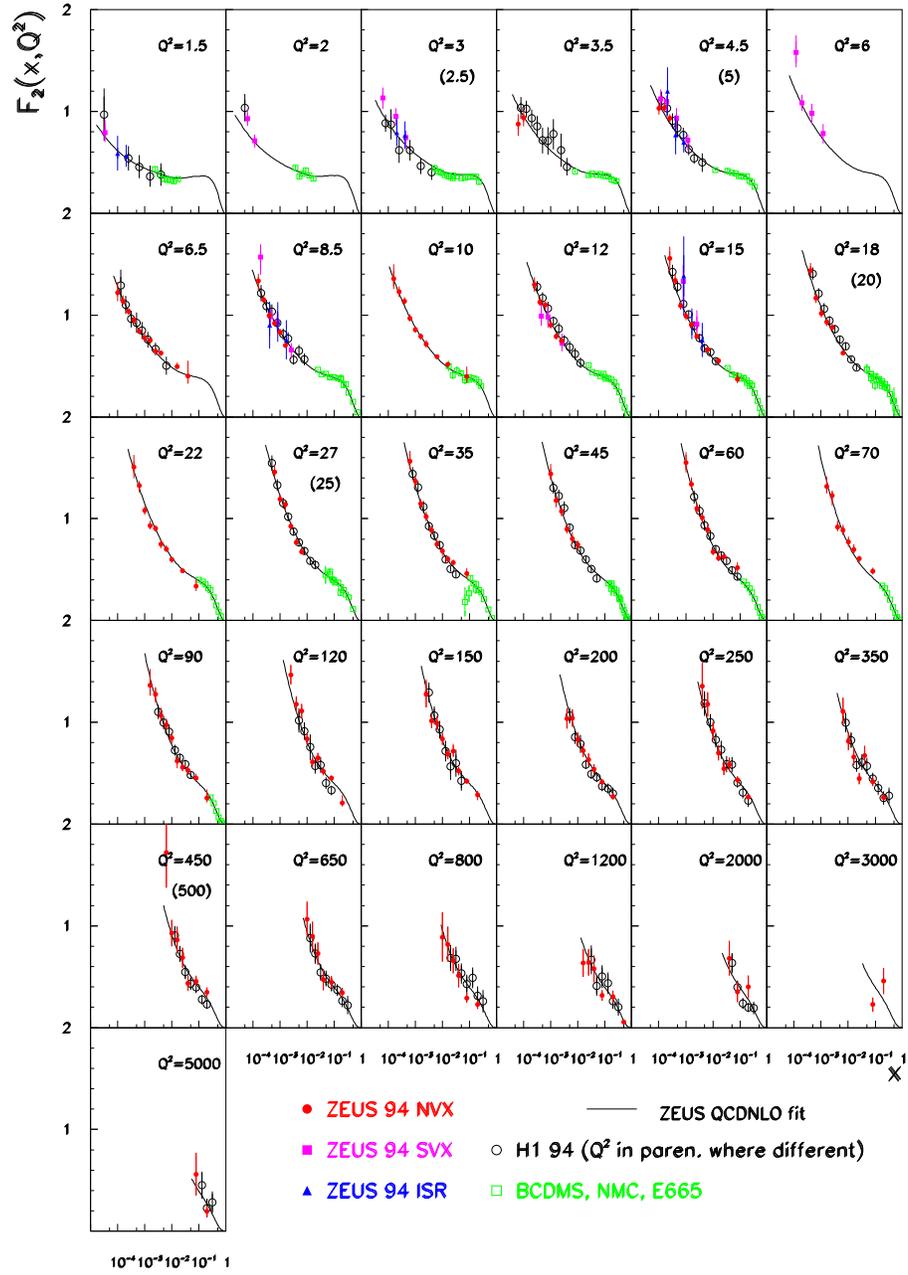}}
    \end{picture}}
\vspace*{0.7cm}
\caption{NLO QCD fit to low-$x$ structure function data for different values 
of $Q^2$.
\label{figherax}}
\end{figure}

The fixed-order approach has been extensively compared with structure function
data during the last few years. The NLO approximation of the DGLAP equations 
describes very well~\cite{grv,mrs,cteq,herafit}$^-$\cite{Ynd} the HERA data
(Fig.~\ref{figherax}), down to low values of $Q^2 \sim 2 \,
{\rm GeV^2}$. The NLO QCD fits simply require slightly steep input parton 
densities at these low momentum scales.

From a phenomenological viewpoint, one may conclude that there is no need for 
BFKL-type corrections to scaling violations in the HERA kinematic region. The
main reason for this is that the LL power-like behaviour of the BFKL gluon
density is valid in the asymptotic regime and the approach to asymptotia is
much delayed~\cite{EKL}$^-$\cite{FRT} by cancellations of logarithmic
corrections that occur at the first perturbative orders in the gluon splitting
function $P_{gg}$.   

\begin{figure}
  \centerline{
    \setlength{\unitlength}{1cm}
    \begin{picture}(0,6)
       \put(0,0){\includegraphics{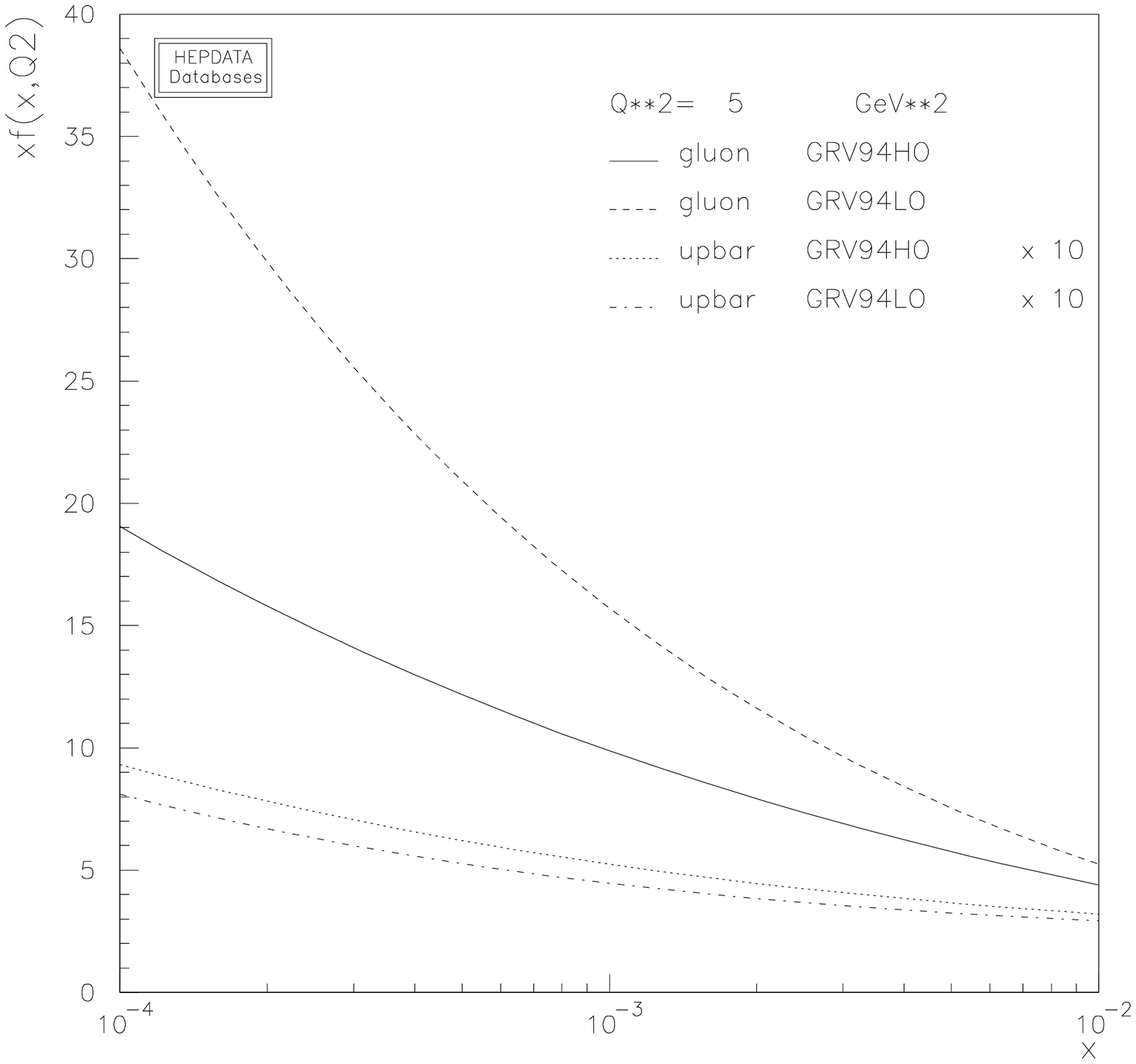}}
    \end{picture}}
\vspace*{0.7cm}    
\caption{Comparison between the LO (GRV94LO) and NLO (GRV94HO) GRV 
parametrizations of the gluon and sea-quark densities at $Q^2=5$~GeV$^2$.
\label{figgrv}}
\end{figure}

Nonetheless, a better way to estimate the relevance of BFKL-type corrections 
is to quantify
the theoretical uncertainty of the NLO predictions,
for instance by comparing LO and NLO results. 
Doing that, one can thus argue that this uncertainty is
sizeable. This feature has recently been re-emphasized in Ref.~\cite{Ynd},
but it was evident since the still-successful GRV parametrization~\cite{grv} 
of parton densities. Going from LO to NLO, one can obtain stable predictions
for the proton structure function $F_2$, but one has to vary the parton
densities a lot, in particular the gluon. 
As shown in Fig.~\ref{figgrv}, the NLO gluon density sizeably differs from its
LO parametrization, not only in absolute normalization but also in $x$-shape.
This can be understood~\cite{sdis} from the fact that the scaling violation of 
$F_2$ is produced by the convolution $P_{qg} \otimes f_g$ (see the right-hand 
side of Eq.~(\ref{df2})). The quark splitting function $P_{qg}$ behaves as
\beq
\label{pqg}
P_{qg}(\as,x) \simeq \as P_{qg}^{(0)}(x) \left[ 1 + 2.2 \frac{C_A \as}{\pi}
\frac{1}{x} + \dots \right] \;\;,
\eeq
where the LO term $P_{qg}^{(0)}(x)$ is flat at small $x$, whereas the NLO
correction is steep. To obtain a stable evolution of $F_2$, the NLO steepness 
of $P_{qg}$ has to be compensated by a gluon density that is less steep 
at NLO than at LO. 

In the large-$x$ region, there is a well known correlation between
$\as$ and $f_g$. At small $x$, there is an analogous strong correlation
between the $x$-shapes of $P_{qg}$ and $f_g$. In the fixed-order DGLAP analysis
of $F_2$, large NLO perturbative corrections at small $x$ can be balanced by
the extreme flexibility of parton density parametrizations. 

\begin{figure}
  \centerline{
    \setlength{\unitlength}{1cm}
    \begin{picture}(0,10.5)
       \put(0,0){\includegraphics{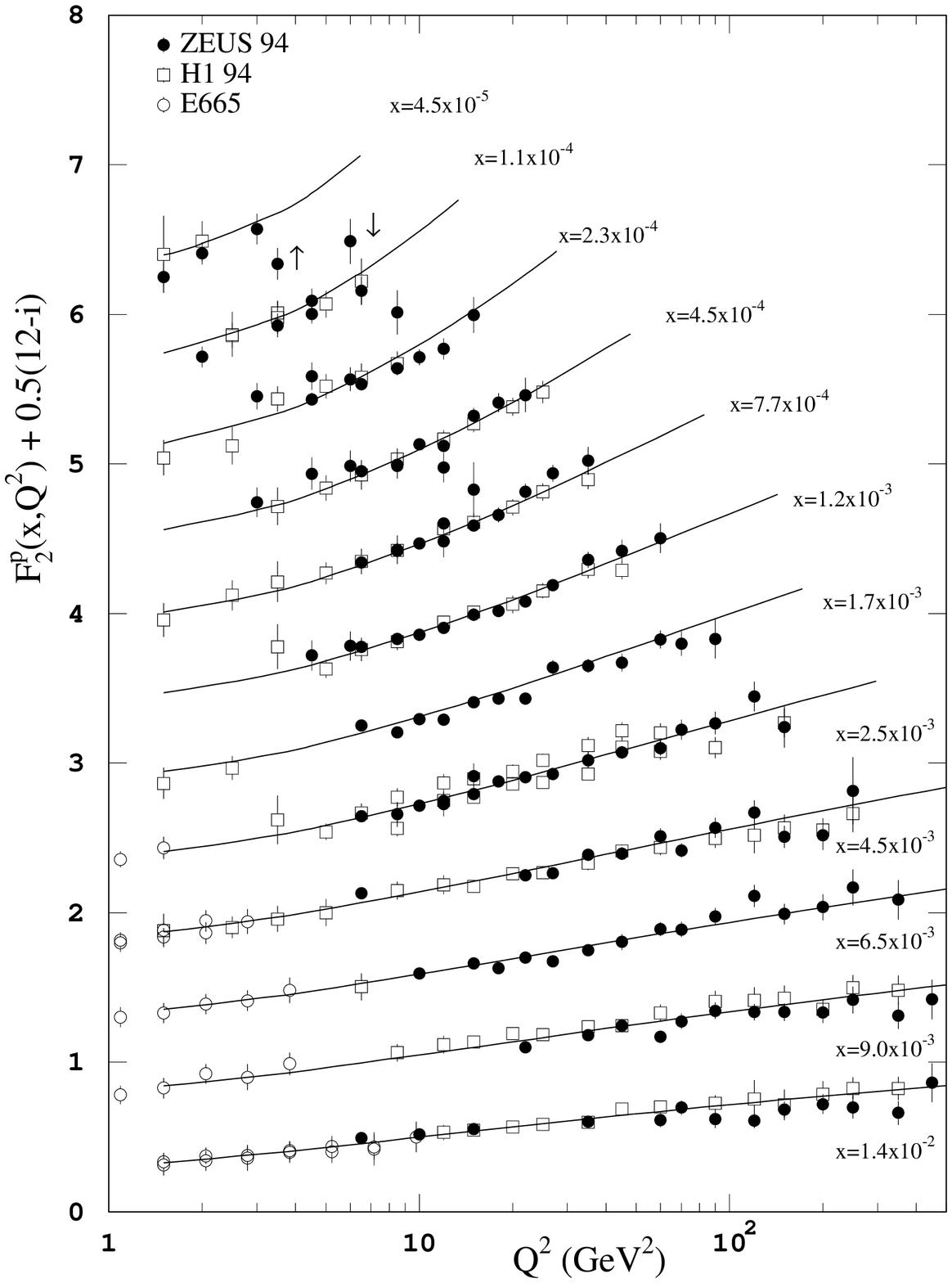}}
    \end{picture}}
\vspace*{-0.7cm}
  \caption{$Q^2$ dependence of the proton structure function $F_2(x,Q^2)$
at various values of $x$. Data from the E665, H1 and ZEUS experiments are
compared with a QCD fit that includes resummation of small-$x$ contributions
to LL accuracy in the gluon channel and to NLL accuracy in the quark channel. 
\label{figthorne} }
\end{figure}

This has to be kept in mind when
concluding on the importance of BFKL dynamics. The NLO steepness
of $P_{qg}$ is the lowest-order manifestation of next-to-leading BFKL 
corrections in the quark channel. Using $k_{\perp}$-factorization~\cite{CCH}
methods, these corrections were resummed to all orders~\cite{CH}
and implemented~\cite{EKL}$^-$\cite{FRT,AKMS}$^-$\cite{thorne} 
in studies of structure functions. The results of a recent fit~\cite{thorne}
are shown in Fig.~\ref{figthorne}.
The present overall picture is that 
small-$x$ resummation leads to a description of the HERA data that is 
comparable to that from NLO analyses, provided that the input gluon density is
less steep than the NLO $f_g$.

More accurate data on $F_2$ are necessary to clarify the phenomenological
relevance of BFKL effects. Measurements of other inclusive observables 
(for instance, the longitudinal structure function $F_L$~\cite{h1fl,thornefl})
can be valuable to disentangle perturbative from non-perturbative dynamics,
that is, to reduce the correlation between $P_{qg}$ and $f_g$ in $F_2$ 
analyses. At the same time, theoretical progress in higher-order calculations 
is very important.

\subsection{BFKL dynamics at NLL accuracy: recent theoretical progress}

The complete evaluation of BFKL contributions to next-to-leading logarithmic 
(NLL) accuracy requires the computation of basic building blocks from
tree-level, one-loop and two-loop amplitudes at high energy. This 
difficult calculational 
program~\cite{FL}$^-$\cite{CCg}, 
led by the work of Fadin and Lipatov, has been completed~\cite{FLsub}.

We are now in a position to start detailed investigations on phenomenological 
and conceptual issues related to the resummation of high-energy logarithms. 

The computation of next-to-leading corrections should permit a quantitative
control of normalizations, scales and 
factorization-scheme~\cite{sdis,q0,BFscheme} uncertainties. 
A preliminary 
evaluation of the NLL corrections to the 
power-like behaviour of the 
gluon density has been performed in Ref.~\cite{CCg}. 
The effective power $\lambda_g$
of the asymptotic $x$-shape of $f_g$ can be written as 
$\lambda_g = 2.65 \;\as(Q) (1 - c \;\as(Q))$ and, 
considering the scale-invariant part of the NLL BFKL contributions,
the correction to $\lambda_g$
is negative and estimated to be quite large~\cite{CCg} $(c \sim 3.5)$.
Negative corrections to $\lambda_g$ are also obtained by 
related investigations~\cite{bmss,LDC} on coherence effects~\cite{coher} and 
the CCFM equation~\cite{coher,CCFM}.

A general theoretical understanding of NLL terms can clarify to what 
extent the BFKL formalism can be used in a purely perturbative 
framework and how it can be matched to the non-perturbative infrared regime.
In the context of structure-function studies, one should thus be able to
precisely identify the kinematic range in $x$ and $Q^2$ where perturbative 
factorization is
valid and not spoiled by $k_{\perp}$-diffusion in the low-momentum 
region~\cite{CC,muel}.

\section{Power corrections}
\label{irpc}
\vspace*{-1mm}
A generic infrared- and collinear-safe observable $R$ that depends on some
large momentum scale $Q$ has the following expression
\beq
\label{rgen}
R(Q) = R_{{\rm pert}}(\as(Q)) + R_{{\rm non-pert}}(Q) \;\;.
\eeq
The term $R_{{\rm pert}}$ denotes the perturbative component that can 
be calculated as power-series expansion in $\as(Q)$ and, thus, behaves as
$(1/\ln Q)^n$. The remaining contribution $R_{{\rm non-pert}}$ is due to
non-perturbative phenomena (hadronization, multiparton scattering, ...)
and is power-behaved, i.e. proportional to $(1/Q)^p$. Since the power $p$ is 
positive, $R_{{\rm non-pert}}$ is suppressed when $Q \to \infty$
but it can be quantitatively relevant at finite values of $Q$.

We have precise theoretical information on $R_{{\rm non-pert}}$ only for
the processes in which OPEs are valid~\cite{ope}. In these cases, one can
write~\cite{muelrev}
\beq
\label{rope}
R_{{\rm non-pert}}(Q) \sim \sum_{p \geq 2} C_p \left( \frac{1}{Q} \right)^p 
\; \langle O_p \rangle 
\;\;,
\eeq
where $\langle O_p \rangle$ denotes the vacuum expectation value of some local 
operator $O_p$, obtained by the basic quark and gluon fields. Note that in
Eq.~(\ref{rope}) we have $p \geq 2$, because in the theory there is no 
(quasi-)local operator with dimension smaller than 2. Note also that
OPEs apply to few quantities such as the total cross section
in $\ee$ annihilation $(p \geq 4)$ and the DIS structure functions $(p \geq 2)$.
For all the other quantities, power-correction contributions are usually 
estimated by using phenomenological approaches (e.g. hadronization models of
Monte Carlo event generators in $\ee$ annihilation).

It is evident that we need a better general understanding of power-suppressed
contributions. On theoretical grounds, one would like to justify power
corrections with $p < 2$, which are experimentally observed in $\ee$ data on 
event shapes (see Fig.~\ref{figdelphimc}), and possibly evaluate their size. 
  
A handle on $R_{{\rm non-pert}}$ can be provided by the study of  
$R_{{\rm pert}}$. It is known that the $\as$-series for the perturbative 
component
$R_{{\rm pert}}$ is not a convergent series but rather an asymptotic
expansion~\cite{muelrev}. This implies that it is not defined in an unambiguous
way. Since $R$ is a physical observable, any ambiguity in 
$R_{{\rm pert}}$ has to
be cancelled by a corresponding ambiguity in $R_{{\rm non-pert}}$. 
Examining the series for $R_{{\rm pert}}$ at large perturbative orders,
one can thus extract some information on $R_{{\rm non-pert}}$.

Two known sources of non-convergent behaviour of the perturbative expansion
are instantons and (infrared) renormalons. Roughly speaking, `instantons' 
are related to large perturbative coefficients due to the large number of
Feynman diagrams at high orders. They lead to non-perturbative corrections 
that are strongly power-suppressed, i.e. that have
high values of $p$.  Infrared renormalons
are related to the low-momentum behaviour of the running coupling $\as$ in 
Feynman diagrams. 

\subsection{Infrared renormalons}

Much theoretical activity has recently been devoted to studying infrared
renormalons~\cite{renrev}. 
A comprehensive review of the field is not feasible because
even a simple list of the relevant references cannot fit into 
several pages of these proceedings. I shall limit myself to recalling
few general points and phenomenological results.  

The basic idea of the renormalon approach to power corrections is the following.
The evaluation of the lowest-order perturbative contribution 
to $R(Q)$ amounts to integrating tree-level Feynman diagrams over the 
momentum $k$ exchanged at the elementary QCD vertices (see the left-hand side
of Eq.~(\ref{rpert})). Higher-order contributions to $R(Q)$ are given by
more complicated Feynman diagrams, including those that produce the
running of the coupling constant $\as$. Therefore, one can approximate
the effects of higher orders by the replacement $\as \to \as(k)$ in the
lowest-order term, as in Eq.~(\ref{rpert}): 
\beq
\label{rpert}  
R_{{\rm pert}}(\as(Q)) \sim \as \int_0^Q \frac{dk}{Q} 
\left(\frac{k}{Q}\right)^p + \dots \;\;\;\; 
\higher \;\;\;\; \;\;\;
\int_0^Q \frac{dk}{Q} \left(\frac{k}{Q}\right)^p \as(k) \;\;.
\eeq
The perturbative series $R_{{\rm pert}}(\as) = \sum_n R_n \as^n$ is then
obtained by inserting the QCD expression for running coupling,
\beq
\as(k) \sim \as(Q)/[1+2 \beta_0 \as(Q) \ln k/Q] = \sum_{n=1}^{\infty}
\as^n(Q) \left( - 2 \beta_0 \ln k/Q \right)^{n-1} \;\;,
\eeq
into the right-hand side of Eq.~(\ref{rpert}) and by integrating term by term.
This leads to perturbative coefficients that grow factorially at high orders:
\beq
\label{rn1}
R_{n+1} = n! \;(1/p) \;(2 \beta_0 /p)^n \;\;.
\eeq
The factorial growth implies that the series for $R_{{\rm pert}}(\as)$ is not 
convergent but can be interpreted as an asymptotic expansion. One should 
truncate the series at the order $n=n_{max}$ at which the ratio of
two successive terms become of order unity and estimate the truncation ambiguity
$\delta R_{{\rm pert}}$ by the size of the last term that is
neglected. The evaluation of $n_{max}$ from Eq.~(\ref{rn1}) gives
\beq
\label{nmax}
( R_{n+1} \as^{n+1}(Q) ) / 
( R_{n} \as^{n}(Q) ) \sim 1 \;\;\;\Longrightarrow \;
n_{max} \sim p/\left(2 \beta_0 \as(Q)\right) \;\;.
\eeq  
Using this value of $n_{max}$ and the QCD expression 
$\as(Q) \sim 1/(2\beta_0 \ln Q/\Lambda_{QCD})$,
one eventually obtains an ambiguity of the perturbative component that has
a power-like behaviour with respect to the large scale $Q$:
\beq
\label{drnp}
\delta R_{{\rm pert}} \sim 
\left( R_{n+1} \as^{n+1}(Q) \right)_{n=n_{max}} \sim e^{-n_{max}} \sim
\Bigl(\Lambda_{QCD}/Q \Bigr)^p \;\;.
\eeq
 
The recipe $\as \to \as(k)$ of Eq.~(\ref{rpert}) to estimate the large-order 
behaviour of the
perturbative series can be justified by summing Feynman graphs in the limit
of a large number $N_f$ of flavours~\cite{bubble} $(N_f \to \infty)$. Further
justifications~\cite{fromres,fromres2} follow from the structure of 
soft-gluon resummation formulae (cf. Eq.~(\ref{r2res})). 

Two main features of the renormalon approach to power corrections are its
predictivity and its non-universality. The power $p$ in Eq.~(\ref{drnp}) is 
exactly equal to that on the right-hand side of Eq.~(\ref{rpert}). Therefore,
the approach unambiguously predicts the type of power correction by relating it
to the dominant low-momentun behaviour of the lowest-order Feynman diagrams
for any given observable. 
Nonetheless, the actual size of the power correction
is not computable unambiguously: the `true' coefficient in front of the factor
$1/Q^p$ is not related in a straightforward and universal way to that obtained 
by evaluating the sole lowest-order Feynman diagrams~\cite{nassey}.

\subsection{Hadronic event shapes in $\ee$ and DIS}

The renormalon predictions on the power behaviour of non-perturbative 
corrections are in agreement~\cite{dispapp} with those from OPEs when the 
latter apply, but can be extended also to other quantities. 

\begin{figure}
  \centerline{
    \setlength{\unitlength}{1cm}
    \begin{picture}(0,8)
       \put(0,0){\includegraphics{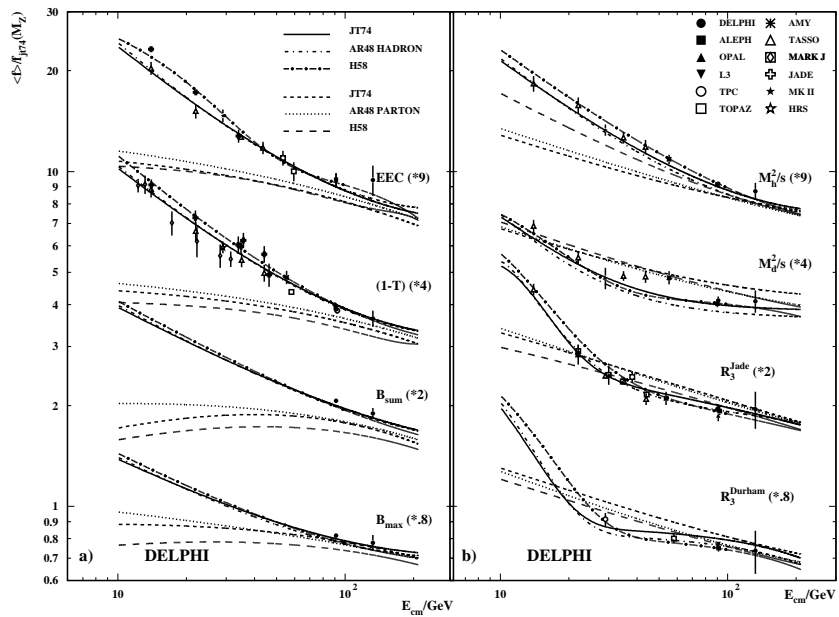}}
    \end{picture}}
\caption{Energy dependence of the mean value of $\ee$ event shapes. The
upper (lower) curves are the predictions of Monte Carlo event generators
at hadron (parton) level.
\label{figdelphimc}}
\end{figure}

Among these quantities, hadronic event shapes in $\ee$ annihilation 
are particularly relevant because their measured mean values receive 
significant non-perturbative contributions of the form $1/Q$. As shown in 
Fig.~\ref{figdelphimc}, parton level predictions of Monte Carlo event 
generators fail
to describe the data, but a good agreement is obtained by taking 
into account $1/Q$ effects produced by the hadronization models that are
built into the Monte Carlo programs. The phenomenological success of these
hadronization models is well known~\cite{hadmodel}, but their relation
with QCD dynamics is still poorly understood on a theoretical basis.

An important result of the renormalon approach is that it 
theoretically predicts~\cite{evren} the 
existence of $1/Q$ non-perturbative corrections to the mean values of event 
shapes. As for the actual size of these corrections, an attempt to overcome
non-universality in renormalon calculations was proposed by 
Dokshitzer and Webber~\cite{DW,dispapp} (DW). The factorial growth of the 
perturbative coefficients in Eq.~(\ref{rn1}) is produced by the 
$k$-integration (see Eq.~(\ref{rpert})) of the perturbative coupling $\as(k)$
down to the Landau pole at $k=\Lambda_{QCD}$. The DW model assumes that a
meaningful non-perturbative definition of $\as(k)$ can be introduced for all
values of $k$. Thus the integral
\beq
\label{as0}
\int_0^{\mu_I} \frac{dk}{\mu_I} \;\as(k) \equiv {\overline \a0}(\mu_I)
\eeq
exists for all $\mu_I \geq 0$ and, using Eq.~(\ref{rpert}) with $p=1$,
the mean value $\langle R(Q) \rangle$ of any
event shape can be written as follows
\beeq
\label{mv}
&&\langle R(Q) \rangle =  \langle R_{\rm pert}(\as(Q)) \rangle_{NLO}
+ \langle R_{\rm non-pert}(Q) \rangle \;\;, \\
\label{mvnp}
&&\langle R_{\rm non-pert}(Q) \rangle = a_R 
\;{\overline \a0}(\mu_I) \;(\mu_I/Q) - a_R  
\left[{\overline \a0}(\mu_I)\right]_{NLO} \;(\mu_I/Q) + \dots
\;\;.
\eeeq
The first term on the right-hand side of Eq.~(\ref{mv}) is the customary
perturbative contribution evaluated up to NLO~\cite{ERT,event}, while the power
correction $\langle R_{\rm non-pert}(Q) \rangle$ is expressed in terms of the
effective non-perturbative parameter ${\overline \a0}(\mu_I)$. To avoid 
double-counting of perturbative contributions, one has to consider
the NLO expansion $\left[{\overline \a0}(\mu_I)\right]_{NLO}$ of the integral
(\ref{as0}) and then subtract the second term
on the right-hand side of Eq.~(\ref{mvnp}).
The coefficient $a_R$ depends on the event shape $R$ and
is obtained by a direct calculation of the corresponding LO Feynman diagrams.

The DW model does not predict the absolute value of 
$\langle R_{\rm non-pert}(Q) \rangle$ 
for each event shape $R$ but parametrizes
all these power corrections in terms of the single `universal' parameter
${\overline \a0}(\mu_I)$ and of calculable process-dependent coefficients $a_R$.
The dots on the right-hand side of Eq.~(\ref{mvnp}) stand for contributions 
that are more power-suppressed (i.e. $(\mu_I/Q)^p$ with $p > 1$) and for 
non-universality corrections to the coefficient $a_R$ of the dominant power. 
These corrections come from higher-order Feynman diagrams~\cite{nassey} and,
in particular, involve integrals of the type 
\beq
\label{abo2}
\int_0^{\mu_I} \frac{dk}{\mu_I} \;\as^2(k) \sim 
{\cal O}({\overline \a0}^2(\mu_I)) \;\;.
\eeq 
If the non-perturbative parameter ${\overline \a0}(\mu_I)$ turns out to be
relatively small, the corrections in Eq.~(\ref{abo2}) can be neglected in a 
first approximation. In phenomenological applications, the infrared matching 
scale $\mu_I$ of Eq.~(\ref{mvnp}) has to be chosen in the range 
$Q \gg \mu_I \gg \Lambda_{QCD}$: $Q \gg \mu_I$, because power corrections of
higher order have to be negligible and $\mu_I \gg \Lambda_{QCD}$, because
$\as(\mu_I)$ still has to be in the perturbative region.

\begin{figure}
  \centerline{
    \setlength{\unitlength}{1cm}
    \begin{picture}(0,7)
       \put(0,0){\includegraphics{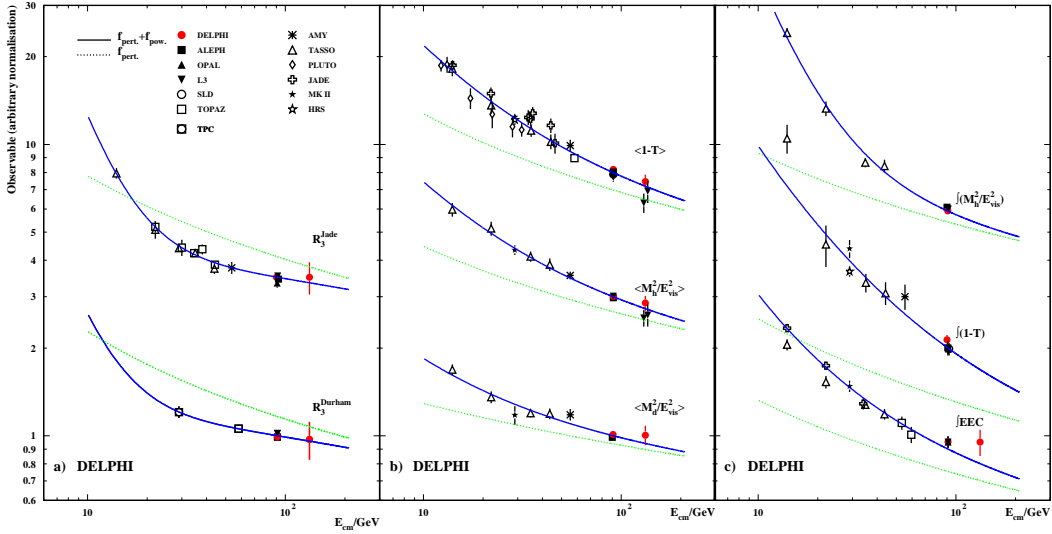}}
    \end{picture}}
\caption{QCD fits for the mean value of $\ee$ event shapes. The solid lines 
correspond to NLO calculations combined with power corrections.
\label{figdelphifit}}
\end{figure}

The DELPHI Collaboration performed a detailed study~\cite{delphipc,wicke}
of the energy dependence of the mean value of $\ee$ event shapes
measured in the centre-of-mass energy range $Q=$ 14--172 GeV.
The data are well described by NLO perturbative calculations~\cite{ERT,event} 
supplemented with power corrections (Fig.~\ref{figdelphifit}). 
Using the predictions
of the DW model for the thrust $T$ and the heavy jet mass $M_h$ 
(Fig.~\ref{figdelphifit}b), a combined fit of $\as(M_Z)$ and the 
non-perturbative parameter ${\overline \a0}(\mu_I)$ gives~\cite{delphipc} 
$\as(M_Z)= 0.116 \pm 0.002 ({\rm exp.}) \pm 0.006 ({\rm th.})$ and
\beq
\label{ab0val}
{\overline \a0}(\mu_I= 2 \,{\rm GeV}) = 
\cases { 0.534 \pm 0.012 \;\;\;\; ({\rm from} \;T) \;,\cr 
0.435 \pm 0.015 \;\;\;\; ({\rm from} \;M_h) \;. \cr}
\eeq
This determination of $\as(M_Z)$ is consistent with those (cf. 
Sect.~\ref{jetratessub})
obtained from analyses of shape distributions in which non-perturbative 
effects are estimated by using Monte Carlo event generators. The value of
${\overline \a0}$ in Eq.~(\ref{ab0val}) suggests that
corrections in the DW model can be kept under control.  
Similar results have been obtained by the JADE~\cite{jaderev}
and ALEPH~\cite{alephpc} Collaborations. 

Note that the fit of Fig.~\ref{figdelphifit}b 
for the jet mass difference $M_d$ corresponds
to an empirical parametrization of the power correction
simply proportional to $1/Q$ rather
than to the predictions of the DW model. In fact, in this case the model gives
a poor quantitative description of the data. This fact signals the presence of
non-universality corrections, which are not included in the naive version of 
the model.

The data in Fig.~\ref{figdelphifit}c are obtained by averaging out 
the event shapes over
a restricted kinematic range~\cite{delphipc} that excludes the two-jet region.
Note that, in the cases of $T$ and $M_h$, the energy dependence of the data
in Fig.~\ref{figdelphifit}c is stronger than 
that in Fig.~\ref{figdelphifit}b. The empirical fits of 
Fig.~\ref{figdelphifit}c 
are consistent with
a dominant power-behaviour of the type $1/Q^2$. This behaviour agrees with that
predicted by calculations of renormalon contributions~\cite{nassey}.

\begin{figure}
  \centerline{
    \setlength{\unitlength}{1cm}
    \begin{picture}(0,7)
       \put(0,0){\includegraphics{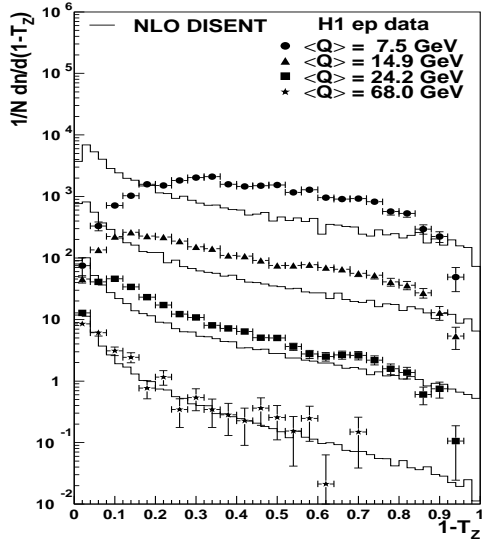}}
    \end{picture}}
\caption{Distribution of the thrust $(T_z)$ of the current fragmentation
region in DIS at different values of $Q$. The histograms are the corresponding
NLO predictions. The spectra are multiplied by factors $10^n, \,n=0,1,2,3$
(downwards).
\label{figh1dist}}
\end{figure}

\begin{figure}
  \centerline{
    \setlength{\unitlength}{1cm}
    \begin{picture}(0,10)
       \put(0,0){\includegraphics{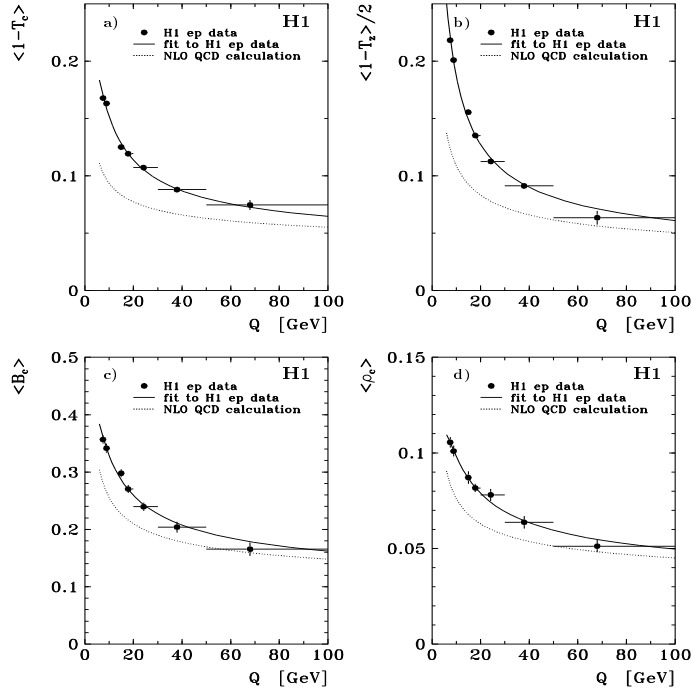}}
    \end{picture}}
\vspace*{-0.8cm}
  \caption{$Q$-dependence of the mean value of DIS event shapes. The solid lines
are QCD fits that include NLO predictions and power corrections according to
the DW model.
\label{figh1av} }
\end{figure}

Lepton--nucleon DIS processes are an ideal place to study the $Q$-dependence
of hadronic observables because one can vary $Q$ over a wide range in a single 
experiment, thus reducing systematic uncertainties. Event shapes in DIS are
analogous to those in $\ee$ annihilation, but they
are usually defined in the current
fragmentation region of the Breit frame. A study~\cite{h1pc} 
of DIS event shapes in the 
momentum region $Q$= 7--100 GeV has recently been performed by
the H1 Collaboration at HERA. As shown in Fig.~\ref{figh1dist}, 
shape distributions
become narrower and jet-like when $Q$ increases. As expected, NLO perturbative
calculations~\cite{mepjet,disent,disaster} describe the data only at 
sufficiently large values of $Q$. At smaller $Q$, non-perturbative contributions
are sizeable. The $Q$-dependence of the mean value of the shape variables is
shown in Fig.~\ref{figh1av}, together with a combined fit that uses 
NLO predictions~\cite{mepjet,disent,disaster} and the DW model for power
corrections~\cite{dispc}. The fit describes the data well and gives~\cite{h1pc}
$\as(M_Z)= 0.118 \pm 0.001 ({\rm exp.}) {}^{+0.007}_{-0.006} ({\rm th.})$ and
${\overline \a0}(\mu_I=2 \,{\rm GeV})= 
0.491 \pm 0.003 ({\rm exp.}){}^{+0.079}_{-0.042} ({\rm th.})$. These values
are consistent with those obtained from $\ee$ event shapes.

Determinations of $\as$ from QCD predictions that include power corrections
are summarized at the end of Table~1. Besides the results from the mean 
values of shape variables, I have also reported a preliminary 
determination~\cite{delphisl} from the longitudinal cross section in $\ee$
annihilation. This determination is based on the NLO calculation in 
Ref.~\cite{nlosl} and on estimates~\cite{slpc} of power-suppressed terms.

At present, it is difficult to quantify the theoretical accuracy of these 
analyses. Universality of power corrections is certainly violated in its
naive form~\cite{DLMS}. In the DW model, the effective non-perturbative 
parameter is
${\overline \a0} \sim 0.5$ and, 
on a phenomenological basis, one may conclude that non-universality effects are
typically smaller than $50\%$. Future investigations on differential
event shape spectra~\cite{wicke,alephpc,DWdist} can produce more definite
quantitative results and give additional insights
into the connection between hadronization and power corrections.

\section{Summary of $\as$}
\label{sumas}

A summary of $\as$ determinations is presented in Table~\ref{tabalphas}.
I shall limit myself to few comments on the comparison
with last year summaries~\cite{Schm,PDG}.

The most relevant new result 
regards a redetermination~\cite{CCFRnew} of $\as$ from $\nu$-nucleon DIS.
The preliminary result, $\as(M_Z)= 0.119 \pm 0.005$, presented by the CCFR
Collaboration at the Warsaw Conference (ICHEP96),
has indeed been confirmed. This result is based on a re-analysis of the CCFR
data, due to a new energy calibration of the detector, and supersedes  
a previous and lower determination $(\as(M_Z) = 0.111 \pm 0.006)$.
The main outcome of this is an increased average value of $\as$ from
DIS measurements. Among DIS determinations, only $\mu$-nucleon data from
BCDMS still prefer~\cite{Roberts} a value of $\as(M_z)$ that is (slightly) 
lower than those from $\ee$ annihilation.

Another new result~\cite{poldis} included in Table~\ref{tabalphas} is from
scaling violation of polarized DIS structure functions. In the case of
polarized DIS processes, a more accurate determination~\cite{Bjsr}
of $\as$ is in principle feasible from the Bjorken sum rule. At present,
however, this determination suffers from uncertainties~\cite{poldis} 
due to the extrapolation of the data in the small-$x$ region, 
where polarized structure functions have not yet been measured.  

The entry from $p{\bar p} \to {\rm jet} + X$ does not refer to an actual
measurement, 
but it illustrates~\cite{ggy} the potential of jet data
from hadron colliders in the determination of $\as$.
Detailed analyses at the Tevatron experiments are in progress.
The updated entry from
$\tau$ decay is from Ref.~\cite{pichrev}. The theoretical uncertainty on
the value~\cite{kobel} from $J/\Psi$ and $\Upsilon$ decays has been 
slightly increased according to estimates~\cite{kobelrev} of additional 
contributions from colour-octet operators.

All the other new or updated entries in Table~\ref{tabalphas} have been
discussed in the previous sections. In particular, the error on $\as$ from
global fits to electroweak observables (cf. Sect.~\ref{nnloas}) includes the
theoretical uncertainty due to the SM assumptions.

\begin{table}[t]
\caption{A summary of measurements of $\as$. 
\label{tabalphas}}
\vspace{0.2cm}
\begin{center}
\footnotesize
\begin{tabular}{|l|c|l|l|c c|c|}
   \hline \large
 & Q & & &  \multicolumn{2}{c|}
{$\Delta \amz $} & Order of \\ 
 Process & [GeV] & $\alpha_s(Q)$ &
  $ \amz$ & exp. & theor. &  perturb. \\
\hline \hline \normalsize
 & & & & & & \\
 GLS sr  
  & 1.73 & \ $0.32 \pm 0.05$ & $0.115 \pm 0.006\ $
  & $\ 0.005\ $ & $\ 0.003\ $ & NNLO \\
 & & & & & & \\ 
 $R_{\tau}$  
    & 1.78 & \ $0.35 \pm 0.04$ & $0.122 \pm 0.005$ & 0.002
    & 0.005 & NNLO \\
 & & & & & & \\
 DIS [polar.]
  & 2.11 & \ $0.31\ ^{+\ 0.08\ }_{-\ 0.06\ }$ & 
  $0.120\ ^{+\ 0.010\ }_{-\ 0.008\ }$
  & $^{+\ 0.004}_{-\ 0.005}$ & $^{+\ 0.009}_{-\ 0.006}$ & NLO \\
 DIS [ HERA $F_2$] 
  & 4.5 & \ $0.23 \pm 0.04$
   & $0.120 \pm 0.010$   &
     0.005 &  0.009 & NLO \\
 DIS [$\nu$] 
  & 5.0 & $0.215 \pm 0.016$
   & $0.119 \pm 0.005$   &
     0.002 &  0.004 & NLO \\
 DIS [$\mu$] 
     & 7.1 & $0.180 \pm 0.014$ & $0.113 \pm 0.005$ & $ 0.003$ &
     $ 0.004$ & NLO \\
 & & & & & & \\
 $b {\bar b}$ mass splitting 
     & 4.1 & $0.223 \pm 0.009 $ & $0.117 \pm \ 0.003\ $
     & 0.000 & 0.003 & LGT \\
 & & & & & & \\
 $\epem$ [$\Upsilon + X$] 
   & 4.1 & $0.228\ ^{+\ 0.045}_{-\ 0.030\ } $ &
   $0.119\ ^{+\ 0.010}_{-\ 0.008\ } $ & $ 0.002 $ 
   & $\ ^{+\ 0.010}_{-\ 0.008\ }$ & NLO \\ 
 $J/\Psi , \Upsilon$ [had. decay]
     & 10.0 & $0.167 \pm 0.020$ & $0.113 \pm 0.010$
     & 0.001 & $ 0.010 $ & NLO \\
 & & & & & & \\
 $\epem$ [$\sigma_{had}$] 
   & 10.5 & \ $0.20 \pm 0.06$ &
   \ $0.13 \pm 0.03$ & $ 0.02 $ & 0.02 & NNLO \\ 
 $\epem$ [ev. shapes] 
   & 22 & $0.161\ ^{+\ 0.016}_{-\ 0.011\ }$ &
   $0.124\ ^{+\ 0.009}_{-\ 0.006\ }$ & 0.005 
   & $\ ^{+\ 0.008}_{-\ 0.003\ }$ & resum. \\ 
 $\epem$ [ev. shapes] 
   & 29 & $0.160\pm 0.012$ &
   $0.131 \pm 0.010$ & 0.006 & 0.008 & resum. \\
 $\epem$ [$\sigma_{had}$] 
   & 34.0 & $0.146\ ^{+\ 0.031}_{-\ 0.026\ }$ &
   $0.124\ ^{+\ 0.021}_{-\ 0.019\ }$ & $\ ^{+\ 0.021}_{-\ 0.019\ }$ 
   & -- & NLO \\
 $\epem$ [ev. shapes] 
   & 35.0 & $0.143\ ^{+\ 0.011}_{-\ 0.007\ }$ &
   $0.122\ ^{+\ 0.008}_{-\ 0.006\ }$ & 0.002 & 
   $\ ^{+\ 0.008}_{-\ 0.005\ }$ & resum. \\
 $\epem$ [ev. shapes] 
   & 44.0 & $0.137\ ^{+\ 0.010}_{-\ 0.007\ }$ &
   $0.122\ ^{+\ 0.008}_{-\ 0.006\ }$ & 0.003 & 
   $\ ^{+\ 0.007}_{-\ 0.005\ }$ & resum. \\
 $\epem$ [ev. shapes] 
   & 58.0 & $0.132\pm 0.008$ &
   $0.123 \pm 0.007$ & 0.003 & 0.007 & resum. \\
 & & & & & & \\
 $pp, p{\bar p} \rightarrow \gamma + X$
  & 4 & $0.206\ ^{+\ 0.042\ }_{-\ 0.024\ }$ & 
  $0.112\ ^{+\ 0.012\ }_{-\ 0.008\ }$
  & 0.006 & $^{+\ 0.010}_{-\ 0.005}$ & NLO \\ 
 $p\bar{p} \rightarrow b\bar{b} + X$ 
    & 20.0 & $0.145\ ^{+\ 0.018\ }_{-\ 0.019\ }$ & $0.113 \pm 0.011 $ 
  & $\ ^{+\ 0.007}_{-\ 0.006}$ & $\ ^{+\ 0.008}_{-\ 0.009}$ 
  & NLO \\
  $p\bar{p} \rightarrow {\rm jet} + X$ 
  & 120 & $0.116 \pm 0.009$ &
  $0.121\pm 0.010$ & 0.008 & 0.005 & NLO \\  
 & & & & & & \\
 $\epem$ [scal. viol.] 
   & 36 & $0.147 \pm 0.014$ &
   $0.125 \pm 0.010$ & 0.006 & 0.008 & NLO \\
 & & & & & & \\
  $\z0$ [e.w. obs.]  
  & 91.2 & $0.120\ ^{+\ 0.005}_{-\ 0.004}$ &
   $0.120\ ^{+\ 0.005}_{-\ 0.004}$ &
  $ 0.003$ & $^{+\ 0.004}_{-\ 0.002}$ & NNLO \\
 $\z0$ [ev. shapes]  
    & 91.2 & $0.122 \pm 0.006$ & $0.122 \pm 0.006$ & $ 0.001$ & $ 0.006$ &
resum. \\
 & & & & & & \\
 $\epem$ [ev. shapes] 
   & 133 & $0.111\pm 0.007$ &
   $0.117 \pm 0.008$ & 0.004 & 0.007 & resum. \\
 $\epem$ [ev. shapes] 
   & 161 & $0.106\pm 0.007$ &
   $0.115 \pm 0.008$ & 0.004 & 0.007 & resum. \\
 $\epem$ [ev. shapes] 
   & 172 & $0.103\pm 0.007$ &
   $0.112 \pm 0.008$ & 0.004 & 0.007 & resum. \\ 
 & & & & & & \\
 DIS [av. ev. shapes] 
   & 7 - 100 & & $0.118\ ^{+\ 0.007\ }_{-\ 0.006\ } $ & 
   0.001 & $ ^{+\ 0.007\ }_{-\ 0.006\ } $ & NLO + p.c. \\
 $\epem$[av. ev. $\!$shapes] 
   & 14 - 172 &  & $0.116\ ^{+\ 0.006\ }_{-\ 0.005\ } $ & 
   0.001 & $ ^{+\ 0.006\ }_{-\ 0.005\ } $ & NLO + p.c. \\
 $\epem$ [$\sigma_L$] 
   & 91.2 & $0.110 \pm 0.016$ &
   $0.110 \pm 0.016$ & 0.013 & 0.010 & NLO + p.c. \\ 
 & & & & & & \\  
\hline
\end{tabular}
\end{center}
\end{table}

\begin{figure}
  \centerline{
    \setlength{\unitlength}{1cm}
    \begin{picture}(0,9)
       \put(0,0){\includegraphics{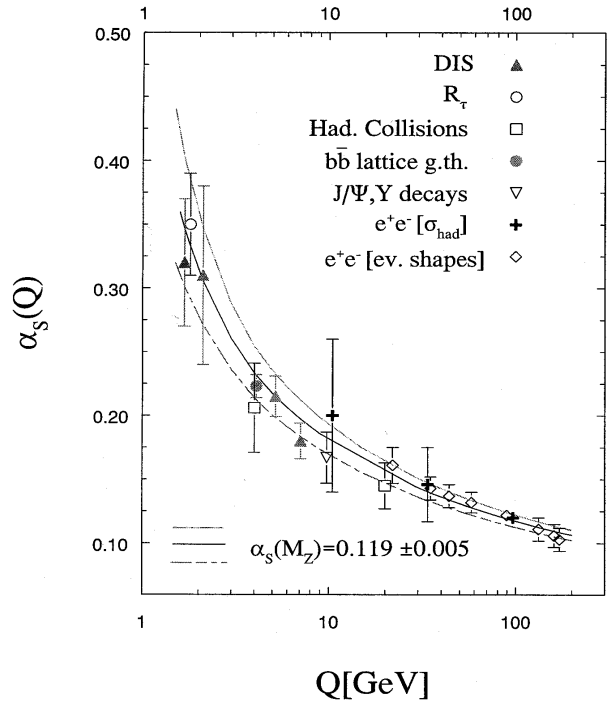}}
    \end{picture}}
\caption{Summary of $\as(Q)$. The lines indicate the QCD predictions in NNLO
for three different values of $\as(M_Z)$. 
\label{figasq}}
\end{figure}

\begin{figure}
  \centerline{
    \setlength{\unitlength}{1cm}
    \begin{picture}(0,9)
       \put(0,0){\includegraphics{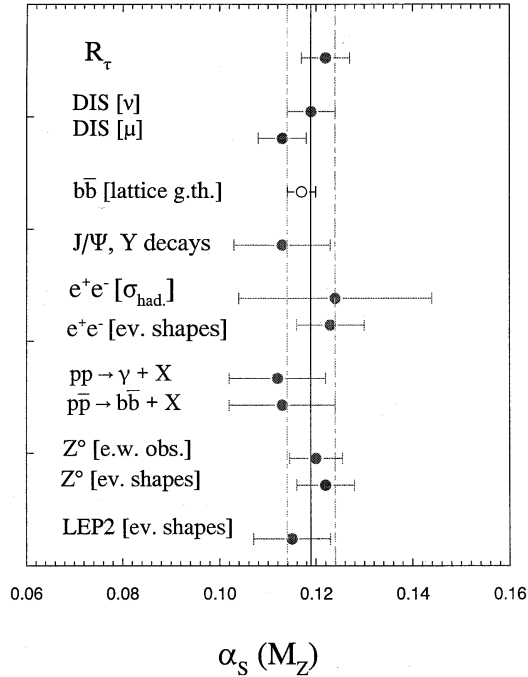}}
    \end{picture}}
\caption{Summary of measurements of $\as(M_Z)$. The band corresponds to the
world average $\as(M_Z)=0.119 \pm 0.005$.
\label{figasmz}}
\end{figure}

The values of $\as$, as a function of the energy scale $Q$ at which they are
measured, are compared with the QCD prediction of a running coupling in 
Fig.~\ref{figasq}. The energy dependence of 
 $\as$ is distinct ($\as(Q)$ varies by a factor of 3) and in
very good agreement with the QCD running over  
two orders of magnitude
in $Q$. It is thus meaningful to evolve all the results to $\as(M_Z)$ according
to perturbative QCD. A significant subset of these values is shown in
Fig.~\ref{figasmz}. The subset is chosen by considering the most relevant 
determinations of $\as$ from each type of process and/or energy range. 
Using these results, my preferred world average determination is
\beq
\label{wavas}
\as(M_Z) = 0.119 \pm 0.005 \;\;.
\eeq
The central value corresponds to the weighted average of the results in
Fig.~\ref{figasmz} (the average does not include the result from 
$b{\bar b}$ mass splitting because of the difficulty in estimating uncertainty
in lattice calculations~\cite{lattice}). The errors on the most precise
determinations of $\as$ are mainly of a theoretical nature and, hence, not
Gaussian and highly correlated among themselves. Because of this reason,
I have not considered any further reduction of the error coming from the
weighting procedure. The uncertainty quoted in Eq.~(\ref{wavas}) is equal to 
the smallest of the errors in Fig.~\ref{figasmz}. 

Different methods,   
used in recent summaries~\cite{Bur,siggi} to quote the world average of 
$\as$, give results similar to that in Eq.~(\ref{wavas}). Less conservative
error estimates ($\Delta \as(M_Z) = 0.003$) are considered in 
Refs.~\cite{Schm,PDG}.

\section{Concluding remarks}
\label{conc}

Owing to high-precision experiments and accurate theoretical calculations,
perturbative QCD is nowadays very well tested in high-energy hadronic
processes.
An oustanding example of this is the accuracy on the
determination of the strong coupling $\as(M_Z)$ and its running. Taken
literally, this accuracy implies that in many processes we can control
strong-interactions dynamics at short distances
to a precision better than $5\%$.
The achieved reliability of perturbative QCD is extremely valuable to estimate
SM backgrounds for new-physics signals, although in some cases the insufficient
knowledge of the non-perturbative parton densities
remains a source of sizeable uncertainty.

Small-$x$ physics deals with a kinematic regime near the borderline between
hard and soft collisions and can lead to a QCD understanding of the
high-energy behaviour of soft hadronic interactions.
In the very near future
substantial progress is expected as a consequence of
the increasing amount of data and of recent theoretical developments in this
field.

The interplay between perturbative and non-perturbative phenomena is one of the
main open issues in QCD. In the past few years, methods inspired by perturbation
theory have been developed to control non-perturbative power corrections to
perturbative predictions. These methods are at present in a stage that is
similar to that of LO perturbative calculations at the end of the 70's.
Forthcoming phenomenological and theoretical studies along these lines 
can play an important role in understanding the hadronization mechanism.
 
\section*{Acknowledgements}
I would like to thank Prof. Albrecht Wagner and Dr. Albert De Roeck for their
excellent organization of the Symposium and Dr. Thomas Gehrmann
for his great help during the Symposium. I am grateful to the many colleagues
who provided me with detailed information about recent results. In particular,
I wish to thank Siggi Bethke, Phil Burrows, Monica Pepe Altarelli
and Heidi Schellman.


\section*{References}
\begin{thebibliography}{99}

\bibitem{Schel}
H.\ Schellman, these proceedings.

\bibitem{Schek}
V.\ Chekelyan, these proceedings.

\bibitem{brull}
A.\ Br\"ull, these proceedings.

\bibitem{SSR}
S.\ Soeldner-Rembold, these proceedings. 

\bibitem{gallo}
E.\ Gallo, these proceedings.

\bibitem{HQtalk}
O.\ Schneider, these proceedings.

\bibitem{Ce}
S.G.\ Gorishny, A.L.\ Kataev and S.A.\ Larin, \pl{259}{144}{91}; L.R.\ 
Surguladze and M.A.\ Samuel, \prl{66}{560}{91}.

\bibitem{Chetyrkin}
K.G.\ Chetyrkin, \pl{391}{402}{97}. 

\bibitem{Passarino}
T.\ Hebbeker, M.\ Martinez, G.\ Passarino and G. Quast, \pl{331}{165}{94}. 

\bibitem{pich}
E.\ Braaten, S.\ Narison and A.\ Pich, \np{373}{581}{92};
F.\ Le Diberder and A.\ Pich, \pl{289}{165}{92}.

\bibitem{neubert}
M.\ Neubert, \np{463}{511}{96}; M.\ Girone and M.\ Neubert, \prl{76}{3061}{96}.

\bibitem{sr}
S.A.\ Larin, F.V.\ Tkachov and J.A.M.\ Vermaseren, \prl{66}{862}{91};
S.A.\ Larin and J.A.M.\ Vermaseren, \pl{259}{345}{91}.

\bibitem{GLSsr}
J.\ Chyla and A.L.\ Kataev, \pl{297}{385}{92}.
 
\bibitem{Bjsr}
J.\ Ellis and M.\ Karliner, \pl{341}{397}{95}; J.\ Ellis, E.\ Gardi, 
M.\ Karliner and M.A.\ Samuel, \pl{366}{268}{96}.

\bibitem{russi}
F.T.\ Tkachov, \pl{100}{65}{81}; K.G.\ Chetyrkin and F.T.\ Tkachov, 
\np{192}{159}{81}.

\bibitem{NNLOaut}
K.G.\ Chetyrkin, \acta{28}{725}{97} and references therein.

\bibitem{ope}
K.\ Wilson, Phys.\ Rev.\ 179 (1969) 1499; K.\ Symanzik, Commun.\ Math.\ Phys.\
32 (1971) 49; C.\ Callan, \pr{5}{3302}{72}; R.\ Brandt, Fortschr.\ Phys.\
18 (1970) 249; M.A.\ Shifman, A.I.\ Vainshtein and V.I.\ Zakharov,
\np{147}{385, 448, 519}{79}.

\bibitem{haidt}
K.\ Hagiwara, D.\ Haidt and S.\ Matsumoto, preprint KEK-TH-512 (hep-ph/9706331).

\bibitem{ewsc}
J.\ Timmermans, these proceedings; D.R.\ Ward, Int. Europhysics Conf. on 
High Energy Physics, EPS 97, Jerusalem, August 1997;
{\it The LEP Electroweak Working Group}, LEPEWWG/97-02.

\bibitem{masscor}
K.G.\ Chetyrkin, A.H.\ Hoang, J.H.\ K\"uhn, M.\ Steinhauser and T.\ Teubner,
preprint DESY 97-25 (hep-ph/9711327) and references therein.

\bibitem{jamin}
M.\ Jamin and A.\ Pich, preprint IFIC-97-06 (hep-ph/9702276).

\bibitem{Vol}
M.B.\ Voloshin, Int.\ J.\ Mod.\ Phys.\ A10 (1995) 2856.

\bibitem{rcleo}
CLEO Coll., R.\ Ammar et al., preprint CLNS-97-1493 (hep-ex/9707018). 

\bibitem{ERT}
R.K.\ Ellis, D.A.\ Ross and A.E.\ Terrano, \np{178}{421}{81};
K.\ Fabricius, I.\ Schmitt, G.\ Kramer and G.\ Schierholz, \zp{11}{315}{81}.

\bibitem{book}
R.K.\ Ellis, W.J.\ Stirling and B.R.\ Webber, {\em QCD and Collider
Physics} (Cambridge University Press, Cambridge, 1996).

\bibitem{GGjet}
W.T.\ Giele and E.W.N.\ Glover, \pr{46}{1980}{92}; 
W.T.\ Giele, E.W.N.\ Glover and D.A.\ Kosower, \np{403}{633}{93}.

\bibitem{FKSjet}
Z.\ Kunszt and D.E.\ Soper, \pr{46}{192}{92};
S.\ Frixione, Z.\ Kunszt and A.\ Signer, \np{467}{399}{96};
S.\ Frixione, preprint ETH-TH-97-14 (hep-ph/9706545).

\bibitem{CSdipole}
S.\ Catani and M.H.\ Seymour, \pl{378}{287}{96}, \np{485}{291}{97}.

\bibitem{mangano}
M.L.\ Mangano and S.J.\ Parke, \prep{200}{301}{91} and references therein.

\bibitem{BDKrev}
Z.\ Bern, L.\ Dixon and D.A.\ Kosower, \ar{46}{109}{96} and references therein.

\bibitem{5part}
Z.\ Bern, L.\ Dixon and D.A. Kosower, \prl{70}{2677}{93}, \np{437}{259}{95};
Z.\ Kunszt, A.\ Signer and Z. Tr\'ocs\'anyi, \pl{336}{529}{94}.

\bibitem{V4part}
E.W.N.\ Glover and D.J.\ Miller, \pl{396}{257}{97};
Z.\ Bern, L.\ Dixon, D.A.\ Kosower and S.\ Weinzierl, \np{489}{3}{97};
J.M.\ Campbell, E.W.N.\ Glover and D.J.\ Miller, preprint DTP-97-44
(hep-ph/9706297);
Z.\ Bern, L.\ Dixon and D.A.\ Kosower, preprint SLAC-PUB-7529
(hep-ph/9708239).

\bibitem{4ee}
A.\ Signer and L.\ Dixon, \prl{78}{811}{97}, \pr{56}{4031}{97};
Z. Nagy and Z.\ Trocsanyi, \prl{79}{3604}{97}, preprint hep-ph/9708343.

\bibitem{3massive} 
G.\ Rodrigo,  Nucl.\ Phys.\ Proc.\ Suppl.\ 54A (1997) 60;
G.\ Rodrigo, A.\ Santamaria and M.\ Bilenkii, \prl{79}{193}{97}; 
W.\ Bernreuther, A.\ Brandenburg and P.\ Uwer, \prl{79}{189}{97};
P.\ Nason and C.\ Oleari, \pl{407}{57}{97}, preprint CERN-TH-97-219
(hep-ph/9709360).

\bibitem{3ppbar}
Z.\ Trocsanyi, \prl{77}{2182}{96}; 
W.B.\ Kilgore and W.T.\ Giele, \pr{55}{7183}{97}.

\bibitem{gluino}
G.R.\ Farrar, preprint RU-97-82 (hep-ph/9710395) and references therein.

\bibitem{delphimb}
DELPHI Coll., S.\ Cabrera et al., Delphi 97-74 Conf 70, submitted to 
this Symposium.

\bibitem{softrev}
See: G.\ Sterman, in {\it Proc. 10th Topical Workshop on Proton-Antiproton 
Collider Physics}, eds. R.\ Raja and J.\ Yoh (AIP Press,
New York, 1996), p.~608; S.\ Catani, preprint LPTHE-ORSAY-97-46 (hep-ph/9709503)
and references therein.
 
\bibitem{CDOTW}
S.\ Catani, Yu.L.\ Dokshitzer, M.\ Olsson, G.\ Tur\-nock and B.R.\ Webber,
\pl{269}{432}{91}.

\bibitem{thrust}
S.\ Catani, G.\ Turnock, B.R.\ Webber and L.\ Trenta\-due, \pl{263}{491}{91}.

\bibitem{hjm}
S.\ Catani, G.\ Turnock and B.R.\ Webber, \pl{272}{368}{91} and 295B (1992)
269.

\bibitem{jetalg}
N.\ Brown and W.J.\ Stirling, \zp{53}{629}{92};
S.\ Bethke, Z.\ Kunszt, D.E.\ Soper and W.J.\ Stirling, \np{370}{310}{92}.

\bibitem{diss}
G.\ Dissertori and  M.\ Schmelling, \pl{361}{167}{95}. 

\bibitem{CTTW}
S.\ Catani, G.\ Turnock, B.R.\ Webber and L.\ Trenta\-due, \np{407}{3}{93}.

\bibitem{event}
Z.\ Kunszt, P.\ Nason, G.\ Marchesini and B.R.\ Webber, in
`Z Physics at LEP 1', CERN 89-08, vol.~1, p.~373.

\bibitem{LEPresum}
S.\ Catani, in {\it Proc. 18th Johns Hopkins Workshop on Current Problems in 
Particle Theory: Theory Meets Experiment}, eds. R.\ Casalbuoni, 
G.\ Domokos, S.\ Kovesi-Domokos and B.\ Monteleoni (World
Scientific, Singapore, 1995), p.~21 and references therein.

\bibitem{pep} 
TPC/Two-Gamma Coll., D.A.\ Bauer et al., preprint LBL-35812 (1994).

\bibitem{tristan}
TOPAZ Coll., Y.\ Ohnishi et al., \pl{313}{475}{93}.

\bibitem{jaderev}
JADE Coll., P.A.\ Movilla Fernandez et al., preprint PITHA-97-27 
(hep-ex/9708034). 

\bibitem{lep2}
L3 Coll., M.\ Acciarri et al., \pl{371}{137}{96}, \pl{404}{390}{97};
OPAL Coll., G.\ Alexander et al., \zp{72}{191}{96}, \zp{75}{193}{97}, 
OPAL-PN281 submitted to this Symposium;
ALEPH Coll., D.\ Buskulic et al., \zp{73}{409}{97}, paper LP-299 submitted
to this Symposium;
DELPHI Coll., J.\ Drees et al., Delphi 97-92 Conf 77, submitted to Int. 
Europhysics Conf. on High Energy Physics, EPS 97, Jerusalem, August 1997.

\bibitem{cdfjet}
CDF Coll., F.\ Abe et al., \prl{77}{438}{96}.

\bibitem{heraq2}
H1 Coll., C.\ Adloff et al., \zp{74}{191}{97};
ZEUS Coll., J.\ Breitweg et al., \zp{74}{207}{97}.

\bibitem{top}
CDF Coll., F.\ Abe et al., \pr{50}{2966}{94}, \prl{74}{2626}{95},
preprint FERMILAB-PUB-97-286-E (hep-ex/9710008);
D0 Coll., S.\ Abachi et al., \prl{74}{2632}{95}, \prl{79}{1203}{97};
P.\ Giromini, these proceedings.

\bibitem{HQnlo}
P.\ Nason, S.\ Dawson and R.K.\ Ellis, \np{303}{607}{88};
W.\ Beenakker, H.\ Kuijf, W.L.\ van Neerven and J.\ Smith, \pr{40}{54}{89}.

\bibitem{sigres}
  E.\ Laenen, J.\ Smith and W.L.\ van Neerven, \np{369}{543}{92}, 
  \pl{321}{254}{94};
  N.\ Kidonakis, J.\ Smith and R.\ Vogt, \pr{56}{1553}{97}.

\bibitem{Berger}
E.L.\ Berger and H.\ Contopanagos, \pl{361}{115}{95}, \pr{54}{3085}{96}.

\bibitem{Bergernew}
E.L.\ Berger and H.\ Contopanagos,  
preprint ANL-HEP-PR-97-01 (hep-ph/9706206).

\bibitem{CMNTtop}
  S.\ Catani, M.L.\ Mangano, P.\ Nason and L.\ Trentadue, \pl{378}{329}{96}.

\bibitem{CMNT}
S.\ Catani, M.L.\ Mangano, P.\ Nason and L.\ Trentadue, \np{478}{273}{96}.

\bibitem{Sterman}    
    G.\ Sterman, \np{281}{310}{87};
    S.\ Catani and L.\ Trentadue, \np{353}{183}{91}, \np{327}{323}{89}. 

\bibitem{kidon}
    N.\ Kidonakis and G. Sterman, \pl{387}{867}{96}, preprint
    EDINBURGH-97-3 (hep-ph/9705234). 

\bibitem{NLOjet}
F.\ Aversa, P.\ Chiappetta, M.\ Greco and J.Ph.\ Guillet, \np{327}{105}{89},
\prl{65}{401}{90};
S.D.\ Ellis, Z.\ Kunszt and  D.E.\ Soper, \pr{40}{2188}{89}, 
\prl{64}{2121}{90};
W.T.\ Giele, E.W.N.\ Glover and D.A.\ Kosower, \np{403}{633}{93}.

\bibitem{tevjet}
F.\ Nang (for the CDF and D0 Collaborations), preprint FERMILAB-CONF-97-192-E,
presented at {\it 32nd Rencontres de Moriond:
QCD and High-Energy Hadronic Interactions}, Les Arcs, March 1997. 

\bibitem{2jets}
CDF Coll., F.\ Abe et al., \prl{77}{5336}{96} (E ibid. 78 (1997) 4307);
D0 Coll., B. Abbot et al., paper LP-197 submitted to this Symposium.

\bibitem{grv}
M.\ Gl\"{u}ck, E.\ Reya and A.\ Vogt, \zp{53}{127}{92}, \zp{67}{433}{95};
A.\ Vogt, \pl{354}{145}{95}.

\bibitem{mrs}
A.D.\ Martin, R.G.\ Roberts and W. J. Stirling, \pl{387}{419}{96}.

\bibitem{cteq}
H.L. Lai et al., \pr{55}{1280}{97}.
 
\bibitem{cteq4hj}
J. Huston et al., \prl{77}{444}{96}.

\bibitem{gual}
G.\ Altarelli, these proceedings.

\bibitem{straub}
B.\ Straub, these proceedings.

\bibitem{lpexer}
S.\ Catani, M.L.\ Mangano and P.\ Nason, unpublished.

\bibitem{CMW}
    S.\ Catani, G. Marchesini and B.R.\ Webber, \np{349}{635}{91};
    H.\ Contopanagos, E.\ Laenen and G. Sterman, \np{484}{303}{97}.

\bibitem{highxunc}
M.A.J.\ Botje (for the ZEUS Coll.), preprint NIKHEF-97-028 (hep-ph/9707289),
presented at {\it 5th Int. Workshop on Deep Inelastic Scattering and QCD} 
(DIS 97), Chicago, April 1997. 
  
\bibitem{tung}
S.\ Kuhlmann, H.L.\ Lai and W.K.\ Tung, preprint MSU-HEP-70316 (hep-ph/9704338).

\bibitem{sxrev}
See: J.\ Kwiecinski, \acta{27}{3455}{96};
H.\ Abramowicz, in {\it Proc. 28th Int. Conference on High-energy Physics} 
(ICHEP 96), eds. Z.\ Ajduk and A.K.\ Wroblewski (World Scientific, 
Singapore, 1997), p.~53
and references therein.

\bibitem{AP}
V.N.\ Gribov and L.N.\ Lipatov, Sov. J. Nucl. Phys. 15 (1972) 438, 
675; G.\ Altarelli and G.\ Parisi,
\np{126}{298}{77}; Yu.L.\ Dokshitzer, Sov. Phys. JETP  46 (1977) 641.

\bibitem{BFKL}
      L.N.\ Lipatov, Sov. J. Nucl. Phys. 23 (1976) 338; E.A.\ Kuraev,
      L.N.\ Lipatov and V.S.\ Fadin, Sov. Phys. JETP  45 (1977) 199; Ya.\
      Balitskii and L.N.\ Lipatov, Sov. J. Nucl. Phys. 28 (1978) 822.

\bibitem{first}
      A.\ De R\'{u}jula, S.L. Glashow, H.D.\ Politzer, S.B.\ Treiman, 
      F.\ Wilczek and A.\ Zee, \pr{10}{1649}{74}.

\bibitem{herafit}
H1 Coll., S.\ Aid et al., \np{470}{3}{96};
ZEUS Coll., M.Derrick et al., \zp{72}{399}{96}. 

\bibitem{bfnlo}
      R.D.\ Ball and S.\ Forte, 
      \pl{335}{77}{94}, 
      \pl{336}{77}{94}, 
      Nucl. Phys. B (Proc. Suppl.) 54A (1997) 163. 
\bibitem{Ynd}
C.\ Lopez, F.\ Barreiro and F.J.\ Yndurain, \zp{72}{561}{96};
K.\ Adel, F.\ Barreiro and F. J. Yndurain, \np{495}{97}{221}.

\bibitem{EKL}
R.K.\ Ellis, Z.\ Kunszt and E.M.\ Levin, \np{420}{517}{94} 
(E \np{433}{498}{95}).

\bibitem{EHW}
       R.K.\ Ellis, F.\ Hautmann and B.R.\ Webber, \pl{348}{582}{95};
       F.\ Hautmann, in {\it QCD and High Energy Hadronic Interactions,
       Proc. 30th Rencontres de Moriond}, ed. J. Tran Thanh Van
       (Editions Fronti\`eres, Gif-sur-Yvette, 1995), p.~133.

\bibitem{BFres}
       R.D.\ Ball and S.\ Forte, \pl{351}{313}{95}, 
      \pl{358}{365}{95}. 

\bibitem{FRT}
J.R.\ Forshaw, R.G.\ Roberts and R.S.\ Thorne, \pl{356}{79}{95}.

\bibitem{sdis}
       S.\ Catani, \zp{70}{263}{96}, \zp{75}{665}{97}.

\bibitem{CCH}
      S. Catani, M. Ciafaloni and F. Hautmann, \pl{242}{97}{90}, 
      \np{366}{135}{91};
      J.C. Collins and R.K. Ellis, \np{360}{3}{91};
      E.M.\ Levin, M.G.\ Ryskin, Yu.M.\ Shabel'skii and A.G.\ Shuvaev, Sov. J.
      Nucl. Phys. 53 (1991) 657.

\bibitem{CH}
S.\ Catani and F.\ Hautmann, \pl{315}{157}{93}, \np{427}{475}{94}.

\bibitem{AKMS}
       A.J.\ Askew, J.\ Kwiecinski, A.D.\ Martin and P.J.\ Sutton, 
       \pr{47}{3775}{93}, \pr{49}{4402}{94};
J.\ Kwiecinski, A.D.\ Martin and A.M.\ Stasto, \pr{56}{3991}{97}.

\bibitem{bfdis96}
S.\ Forte and R.D.\ Ball, in {\it Proc. Int. Workshop on Deep 
Inelastic Scattering and Related Phenomena}, DIS 96, eds. G.\ D'Agostini 
and A. Nigro  (World Scientific, Singapore, 1997), p.~172.

\bibitem{blum}
J.\ Bl\"{u}mlein, S.\ Riemersma and A. Vogt, 
Nucl. Phys. B (Proc. Suppl.) 51C (1996) 30.

\bibitem{thorne}
R.S.\ Thorne, \pl{392}{463}{97}, preprint hep-ph/9710541.  

\bibitem{h1fl}
H1 Coll., C.\ Adloff et al., \pl{393}{452}{97}. 

\bibitem{thornefl}
R.S.\ Thorne, preprint RAL-TR-97-039 (hep-ph/9708302).

\bibitem{FL}
V.S.\ Fadin and L.N.\ Lipatov, 
      \sj{50}{712}{89}, \np{406}{259}{93}, \np{477}{767}{96};
V.S.\ Fadin, R.\ Fiore and A.\ Quartarolo, 
       \pr{50}{5893}{94}, \pr{53}{2729}{96}; 
V.S.\ Fadin, JETP Lett. 61 (1995) 346;
V.S.\ Fadin, R.\ Fiore and M.I. Kotskii, \pl{387}{593}{96};
V.S.\ Fadin, M.I.\ Kotskii and L.N. Lipatov, preprint BUDKERINP-96-92
(hep-ph/9704267). 

\bibitem{CCq}
G.\ Camici and M.\ Ciafaloni, \pl{386}{341}{96}, \np{496}{305}{97}.

\bibitem{DD} 
V.\ Del Duca, \pr{54}{989}{96}, \pr{54}{4474}{96};
V.\ Del Duca and C.R.\ Schmidt, preprint EDINBURGH-97-26 (hep-ph/9711309).     

\bibitem{CCg}
G.\ Camici and M.\ Ciafaloni, preprint hep-ph/9707390; 
M.\ Ciafaloni, preprint hep-ph/9709390, presented at  
{\it Ringberg Workshop on New Trends in HERA Physics}, 
Ringberg Castle, May 1997.

\bibitem{FLsub}
V.S.\ Fadin, M.I.\ Kotskii and L.N. Lipatov, preprint BUDKERINP-97-56, submitted
to this Symposium.
 
\bibitem{q0}
M.\ Cia\-fa\-lo\-ni,  \pl{356}{74}{95}.

\bibitem{BFscheme}
       R.D.\ Ball and S.\ Forte, \pl{359}{362}{95}. 

\bibitem{bmss}
G.\ Bottazzi, G.\ Marchesini, G.P.\ Salam and M.\ Scorletti, 
preprint IFUM-552-FT (hep-ph/9702418).

\bibitem{LDC}
B.\ Andersson, G.\ Gustafson and J.\ Samuelsson, \zp{71}{613}{96},
\np{467}{443}{96}.

\bibitem{coher}
M.\ Ciafaloni, \np{296}{49}{88}. 

\bibitem{CCFM}
S.\ Catani, F. Fiorani and G. Marchesini, \pl{234}{339}{90}, \np{336}{18}{90}. 

\bibitem{CC}
G.\ Camici and M.\ Ciafaloni, \pl{395}{118}{97}.
 
\bibitem{muel}
A.H. Mueller, \pl{396}{251}{97}.

\bibitem{muelrev}
A.H. Mueller, in {\it QCD: 20 Years Later}, eds. P.M.\ Zerwas and 
H.A. Kastrup (World Scientific, Singapore, 1993), p.~162
and references therein. 

\bibitem{renrev}
See: R.\ Akhoury and V.I. Zakharov, Nucl.\ Phys.\ B (Proc. Suppl.)
54A (1997) 217; M.\ Beneke, preprint CERN-TH-97-134 (hep-ph/9706457) and 
references therein.

\bibitem{bubble}
P.\ Ball, M.\ Beneke and V.M.\ Braun, \np{452}{563}{95} and references therein.

\bibitem{fromres}
G.P.\ Korchemsky and G.\ Sterman, \np{437}{415}{95};
R.\ Akhoury and V.I.\ Zakharov, \pl{357}{646}{95}, \np{465}{295}{96}.

\bibitem{fromres2} 
M.\ Beneke and V.M.\ Braun, \np{454}{253}{95}.

\bibitem{nassey}
P.\ Nason and M.H.\ Seymour, \np{454}{291}{95}.

\bibitem{dispapp}
Yu.L.\ Dokshitzer, G.\ Marchesini and B.R.\ Webber, \np{469}{93}{96}.

\bibitem{hadmodel}
I.G.\ Knowles and G.D.\ Lafferty, J.\ Phys.\ G23 (1997) 731
and references therein.

\bibitem{evren}
B.R.\ Webber, \pl{339}{148}{94};
A.V.\ Manohar and M.B.\ Wise, \pl{344}{407}{95}.

\bibitem{DW}
Yu.L.\ Dokshitzer and B.R.\ Webber, \pl{352}{451}{95}.

\bibitem{delphipc}
DELPHI Coll., P.\ Abreu et al., \zp{73}{243}{97}.

\bibitem{wicke}
D.\ Wicke, preprint hep-ph/9708467, presented at 
{\it Int. Euroconference on Quantum Chromodynamics}, QCD 97, Montpellier,
July 1997.

\bibitem{alephpc}
ALEPH Coll., paper LP-258 submitted to this Symposium.

\bibitem{h1pc}
H1 Coll., C.\ Adloff et al., \pl{406}{256}{97}.

\bibitem{mepjet}
E.\ Mirkes and D.\ Zeppenfeld, \pl{380}{205}{96}, 
preprint MADPH-97-1002 (hep-ph/9706437).

\bibitem{disent}
S.\ Catani and M.H.\ Seymour, \np{485}{291}{97}, in {\it Proc.
Workshop on Future Physics at HERA}, eds. G.\ Ingelman, A.\ De Roeck and 
R.\ Klanner (DESY, Hamburg, 1996), p.~519.

\bibitem{disaster}
D.\ Graudenz, preprint PSI-PR-97-20 (hep-ph/9708362), 
preprint hep-ph/9710244.
 
\bibitem{dispc}
M.\ Dasgupta and B.R.\ Webber, preprint CAVENDISH-HEP-96-5 (hep-ph/9704297).

\bibitem{delphisl}
DELPHI Coll., P.\ Abreu et al., preprint DELPHI 97-69 CONF 55, submitted to
this Symposium.

\bibitem{nlosl}
P.J.\ Rijken and W.L.\ van Neerven, \pl{386}{422}{96},
\np{487}{233}{97}.

\bibitem{slpc}
M.\ Dasgupta and B.R.\ Webber, \np{484}{247}{97};
M.\ Beneke, V.M.\ Braun and L.\ Magnea, \np{497}{297}{97}.

\bibitem{DLMS}
Yu.L.\ Dokshitzer, A.\ Lucenti, G.\ Marchesini and G.P. Salam,
preprint  IFUM-573-FT (hep-ph/9707532). 

\bibitem{DWdist}
Yu.L. Dokshitzer and B.R. Webber, \pl{404}{321}{97}.

\bibitem{Schm}
M.\ Schmelling, in {\it Proc. 28th Int. Conference on High-energy Physics}
(ICHEP 96), eds. Z.\ Ajduk and A.K.\ Wroblewski (World Scientific, 
Singapore, 1997), p.~91.

\bibitem{PDG}
Particle Data Group, Review of Particles Properties, R.M.\ Barnett et al.,
\pr{54}{1}{96}.

\bibitem{CCFRnew}
CCFR Coll., W.G.\ Seligman et al., preprint NEVIS-REPORT-292 (hep-ex/9701017).

\bibitem{Roberts}
R.G.\ Roberts, preprint RAL-TR-97-024 (hep-ph/9706269), presented at 
{\it 5th Int. Workshop on Deep Inelastic Scattering and QCD} (DIS 97), 
Chicago, April 1997. 
 
\bibitem{poldis}
G.\ Altarelli, R.D.\ Ball, S.\ Forte and G.\ Ridolfi, \np{496}{337}{97}.

\bibitem{ggy}
W.T.\ Giele, E.W.N.\ Glover and J.\ Yu, \pr{53}{120}{96}.

\bibitem{pichrev}
A.\ Pich, preprint FTUV-97-22 (hep-ph/9704453), to appear in 
{\it Heavy Flavors II}, 
eds. A.J.\ Buras and M.\ Lindner (World Scientific, Singapore, 1997). 

\bibitem{kobel}
M.\ Kobel, in {\it Perturbative QCD and Hadronic Interactions, Proc.
27th Rencontres de Moriond}, ed. J. Tran Thanh Van
(Editions Fronti\`eres, Gif-sur-Yvette, 1992), p.~145.

\bibitem{kobelrev}
M.\ Gremm and A.\ Kapustin, \pl{407}{323}{97}.

\bibitem{lattice}
M.\ L\"uscher, these proceedings.

\bibitem{Bur}
P.N.\ Burrows, preprint SLAC-PUB-7631 (hep-ex/9709010), presented at
{\it  17th Int. Conference on Physics in Collision} (PIC 97),
Bristol, June 1997. 

\bibitem{siggi}
S.\ Bethke, preprint PITHA-97-37 (hep-ex/9710030), presented at {\it 
Int. Euroconference on Quantum Chromodynamics}, QCD 97, 
Montpellier, July 1997. 
 



\end{thebibliography}
\end{document}